\documentclass[prx, twocolumn, nofootinbib, showpacs, amsmath, amssymb, floatfix, eqsecnum]{revtex4-1}
\usepackage{amsmath}
\usepackage{amssymb}
\usepackage{amsthm}
\usepackage{amsfonts}
\usepackage{comment}
\usepackage{xcolor}
\usepackage{cancel}

\usepackage[normalem]{ulem}
\usepackage{graphicx,color,framed,caption}
\usepackage{hyperref}
\usepackage{times}
\usepackage{enumerate}
\usepackage{lipsum}
\usepackage{slashed}
\usepackage{url}
\usepackage{bbm}
\usepackage{wasysym}
\hypersetup{
    colorlinks=true, 
    linktoc=all,     
    linkcolor=black,  
}

\def \beq {\begin{equation}}
\def \eeq {\end{equation}}
\def \beqa {\begin{eqnarray}}
\def \eeqa {\end{eqnarray}}
\def \bseq {\begin{subequations}}
\def \eseq {\end{subequations}}

\begin{document}

\title{Anomalies of $(1+1)D$ categorical symmetries}
\author{Carolyn Zhang}
\author{Clay C\'{o}rdova}
\affiliation{Department of Physics, Kadanoff Center for Theoretical Physics, University of Chicago, Chicago, Illinois 60637,  USA}
\date{\today}

\begin{abstract}
We present a general approach for detecting when a fusion category symmetry is anomalous, based on the existence of a special kind of Lagrangian algebra of the corresponding Drinfeld center. The Drinfeld center of a fusion category $\mathcal{A}$ describes a $(2+1)D$ topological order whose gapped boundaries enumerate all $(1+1)D$ gapped phases with the fusion category symmetry, which may be spontaneously broken. There always exists a gapped boundary, given by the \emph{electric} Lagrangian algebra, that describes a phase with $\mathcal{A}$ fully spontaneously broken. The symmetry defects of this boundary can be identified with the objects in $\mathcal{A}$. We observe that if there exists a different gapped boundary, given by a \emph{magnetic} Lagrangian algebra, then there exists a gapped phase where $\mathcal{A}$ is not spontaneously broken at all, which means that $\mathcal{A}$ is not anomalous. In certain cases, we show that requiring the existence of such a magnetic Lagrangian algebra leads to highly computable obstructions to $\mathcal{A}$ being anomaly-free. As an application, we consider the Drinfeld centers of $\mathbb{Z}_N\times\mathbb{Z}_N$ Tambara-Yamagami fusion categories and recover known results from the study of fiber functors.
\end{abstract}

\maketitle

\makeatletter
\def\l@subsection#1#2{}
\def\l@subsubsection#1#2{}
\makeatother
\tableofcontents 

\section{Introduction}\label{sintro}

Symmetry plays a crucial role in the modern understanding of quantum field theory and renormalization group flows. A microscopic model invariant under a given set of symmetry transformations gives rise to a macroscopic theory which must also realize the same symmetry operators.  In particular, 't Hooft anomalies, which may be broadly understood as obstructions to gauging a given symmetry, are quantized invariants of the symmetry action.  As such, they are calculable at short distances and scale invariant, and hence are powerful tools for constraining dynamics.  Specifically, a non-vanishing 't Hooft anomaly obstructs the existence of a trivially gapped realization of the symmetry.\footnote{A trivially gapped theory is one with a unique vaccuum on any spatial manifold.  Such a system is also known as a symmetry protected topological order (SPT) or an invertible quantum field theory.}  

A familiar class of examples, that we aim to generalize below, are $(1+1)D$ theories invariant under a discrete symmetry group $G$. (In the terminology of \cite{gaiotto2015}, these are 0-form symmetries.)  Concretely, one may consider such symmetries as defined by operators acting on a Hilbert space with composition described by group multiplication.  Alternatively, from the viewpoint of field theory, the symmetries are characterized by line operators with topological correlation functions. There is one such defect for each $g\in G$ and their fusion encodes the group law. In a spontaneously broken realization of the symmetry, such defects flow to domain walls separating distinct vacua.  

From the symmetry operators/defects one may extract an $F$ symbol. (For an explicit construction see \cite{else2014, kawagoe2021}.)  This is a map $F: G\times G\times G\to U(1)$ that describes associativity of fusion of defects of the symmetry.  When the $F$ symbol is a nontrivial 3-cocycle, the symmetry has a 't Hooft anomaly. Therefore, for bosonic systems with a unitary symmetry $G$, the different equivalence classes of 't Hooft anomalies are classified by $H^{3}(G,U(1))$ \cite{chen2013, Kapustin:2014zva}.  This classification also has an elementary interpretation via anomaly inflow \cite{Dijkgraaf:1989pz, Callan:1984sa, chen2013, Kapustin:2014zva}.  The anomalous $(1+1)D$ system with symmetry $G$ may be realized as the edge of a $(2+1)D$ invertible phase with classical background $G$  gauge fields and action defined by the cohomology class $F\in H^{3}(G,U(1))$.  Alternatively, one can also find the same anomalous $(1+1)D$ system at the edge of a topological gauge theory defined by dynamical $G$ gauge fields and action $F\in H^{3}(G,U(1))$.  In this case, the bulk $(2+1)D$ theory resides on a slab geometry where one end supports the anomalous $(1+1)D$ system and the other end supports a canonical Dirichlet boundary condition.  This gives the simplest example of the paradigm of symmetry topological field theories \cite{Fuchs:2002cm, Freed:2012bs, Gaiotto:2020iye, PhysRevResearch.2.033417, Apruzzi:2021nmk, Freed:2022qnc, Kaidi:2022cpf, Chatterjee:2022tyg, vanBeest:2022fss, Freed:2022iao, Kaidi:2023maf}, a bulk $(2+1)D$ TQFT characterizing a symmetry and anomaly, which will feature throughout this work.

Symmetries characterized by a group describe a small subset of more general fusion category symmetries \cite{bhardwaj2018,chang2019,thorngrenwang1,thorngrenwang2}.  At the operator level, a fusion category symmetry is described by a Hilbert space and a set of global operators that form an algebra rather than a group. In a field theory, these are topological line operators with general fusion algebras. Importantly, a global symmetry operator may not have an inverse, and for this reason such symmetries are often referred to as non-invertible.  A simple example of this kind of symmetry is the Ising fusion category symmetry, found at the critical point of an Ising spin chain. It is defined by three topological lines $\{\mathbf{1},\psi,\sigma\}$, where $\psi$ is the $\mathbb{Z}_2$ Ising symmetry and $\sigma$ is the Kramers-Wannier duality line, present because the Ising model at the critical point is duality invariant.  Their fusion algebra is:
\begin{equation}\label{isingf}
\psi\times\psi=\mathbf{1}\qquad \sigma\times\sigma=\mathbf{1}+\psi.
\end{equation}
Note that $\sigma$ does not have an inverse, because there is no operator that fuses with $\sigma$ to give only $\mathbf{1}$. The Ising example \eqref{isingf} also illustrates the key idea of an anomalous fusion category symmetry, namely as we discuss below, there is no SPT phase realizing the symmetry algebra above.  Indeed, the duality line $\sigma$ may be either spontaneously broken leading to a three-fold ground state degeneracy, or preserved necessitating a gapless conformal field theory at long distances.  

Beyond the critical Ising model, non-invertible symmetries characterized by fusion categories are ubiquitous in conformal field theories \cite{Verlinde:1988sn, Petkova:2000ip, Frohlich:2004ef, Frohlich:2006ch, Frohlich:2009gb, chang2019, thorngrenwang1,thorngrenwang2}, gapped boundaries of $(2+1)D$ topological field theories \cite{lan2015,albert2021,freed2022,kaidi2022}, and anyon chains \cite{feiguin2007, buican2017, Huang:2021nvb, Vanhove:2021zop, inamura2022}.  Recently, they have featured in diverse physical applications, including constraints on the operator spectrum of CFTs \cite{Lin:2022dhv, Lin:2023uvm}, and renormalization group flows in gauge theories \cite{Komargodski:2020mxz}.  Higher-dimensional analogs of these symmetries have also been recently been constructed in \cite{Kaidi:2021xfk, choi2022} and similarly used to provide insight into dynamics \cite{choi2022, Choi:2022zal, apte2022, Kaidi:2023maf}.

As in the case of invertible (group-like) symmetries, a key question is to understand the possible phases that can support a given fusion category symmetry $\mathcal{A}$.  Both gapless symmetry preserving phases and gapped spontaneous symmetry breaking phases realizing $\mathcal{A}$ are always possible.  By contrast, the existence of an SPT phase realizing $\mathcal{A}$ is delicate.  By analogy with the case of invertible symmetry, when an SPT realizing $\mathcal{A}$ exists we say that $\mathcal{A}$ is non-anomalous, while if no such SPT phase exists, we say that $\mathcal{A}$ is anomalous.  Characterizing anomalies of a given fusion category $\mathcal{A}$ is the main aim of this work.

\subsection{Fusion categories and their anomalies}

To frame our discussion, let us briefly review the defining data of a fusion category $\mathcal{A}$.  For a more detailed treatment see e.g. \cite{kitaev2006,etingof2016,barkeshli2019}. 

\subsubsection{Fusion rules}
The fusion category $\mathcal{A}$ is described by a set of simple objects $\{a\}$, and their fusion rules 
\begin{equation}\label{fusioncat}
a\times b=\sum_cN_{ab}^{c} c.
\end{equation}
The fusion coefficients $N_{ab}^c$ are non-negative integers which specify the number of different ways that $a$ and $b$ can fuse into $c$. The set of simple objects must include a unique vacuum object which we label $\mathbf{1}$, with $N_{a\mathbf{1}}^c=N_{\mathbf{1}a}^c=\delta_{a,c}$ for all $a\in\mathcal{A}$. We also require every simple object $a$ to have a unique ``antiparticle" $\bar{a}$, such that $\mathbf{1}$ is included in their fusion product: 
\begin{equation}
    a\times\bar{a}=\mathbf{1}+\cdots.
\end{equation}
Note that it is possible that $a=\bar{a}$.  The fusion rules are required to be associative hence:
\begin{equation}
\sum_eN_{ab}^eN_{ec}^d=\sum_fN_{af}^dN_{bc}^f.
\end{equation}
Each simple object has a real, positive quantum dimension $d_a$, corresponding to the following process:
\begin{equation}
\includegraphics{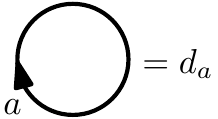},
\end{equation}
and \eqref{fusioncat} constrains these quantum dimensions to satisfy
\begin{equation} \label{fusionds}
d_ad_b=\sum_cN_{ab}^cd_c.
\end{equation}

Although they do not completely define the fusion category, we note that the fusion rules may already be anomalous.  In other words, a given set of fusion rules may be incompatible with a trivially gapped realization of the fusion category $\mathcal{A}$.  Specifically, a necessary (but as we see below not in general sufficient) condition for a trivially gapped realization of $\mathcal{A}$ is that each quantum dimension $d_{a}$ must be an integer. 

A simple argument for this was given in \cite{chang2019}. In a trivially gapped phase, there is a unique ground state in the Hilbert space of the theory quantized on a circle.  Alternatively in radial quantization, this state is dual to the unique unit operator.  The quantum dimension of an object $a$ then gives the vacuum expectation value of the associated topological line operator:
\begin{equation}
    d_{a}=\langle 0|a|0 \rangle.
\end{equation}
Equivalently, since there is only one state we can view the above as a torus partition function with an insertion of the line $a$.  Using modular invariance, we can alternatively interpret this same quantity as the trace of the Hilbert space $H_{a}$ of states on $S^{1}$ with $a$ inserted at a point in space and extending in time.  By the state operator map, this is the same as looking at point operators which can end the line $a$.  Therefore we find
\begin{equation}\label{qdanom}
    d_{a}=\mathrm{Tr}_{H_{a}} \in \mathbb{N},
\end{equation}
where the last statement follows from the fact that the trace simple computes the dimension of the Hilbert space $H_{a}$. 

Note that the constraint \eqref{qdanom} is trivially satisfied by all invertible fusion categories.  By contrast, an example where this constraint is not satisfied is the Ising fusion category \eqref{isingf}, which has $d_{\sigma}=\sqrt{2}$. (\ref{qdanom}) thus gives a simple proof that this fusion category cannot be realized in a trivially gapped phase.\footnote{For higher dimensional analogs of this constraint, see \cite{Cordova:2019bsd, Cordova:2019jqi}.}  

\subsubsection{$F$ symbols}

\begin{figure}[tb]
   \centering
   \includegraphics{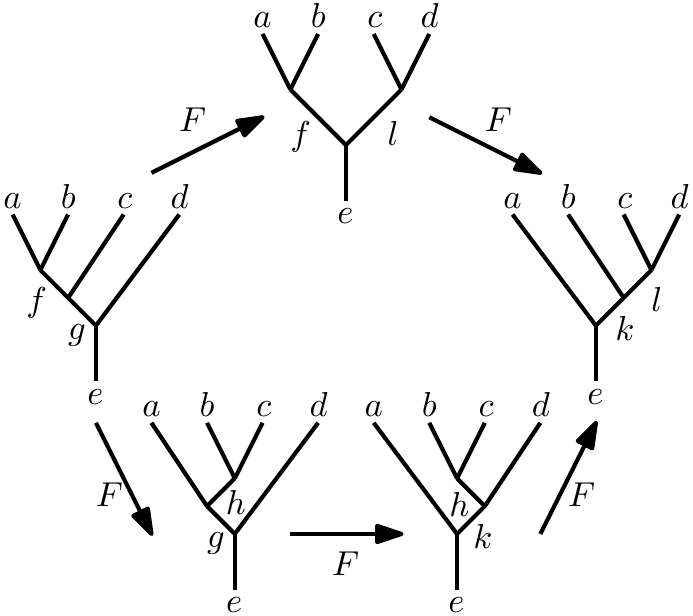} 
   \caption{The pentagon equation is a consistency equation ensuring the two paths to the same configuration given by the top part of the pentagon and the bottom part of the pentagon match. This puts strict constraints on the allowed $F$ symbols. For fusion categories describing invertible symmetries, the pentagon equation is equivalent to the cocycle condition on the $F$ symbol. Note that we have omitted the arrows on the anyon lines for clarity of the figure.}
   \label{fig:pentagon}
\end{figure}

Beyond the algebra \eqref{fusioncat}, the fusion category $\mathcal{A}$ is also equipped with an $F$ symbol that describes the associativity of fusion: 
\begin{equation}
\includegraphics{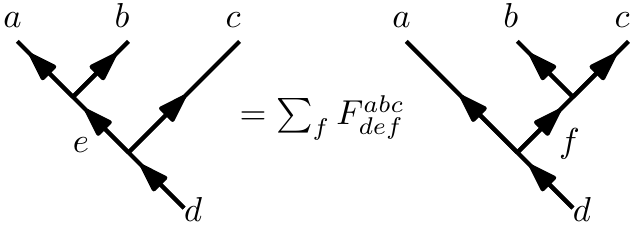}.
\end{equation}
(In this work, we will assume that $N_{ab}^c\in\{0,1\}$, so the $F$ symbol does not have additional indices, but the generalization is straightforward. Because we assume that $N_{ab}^c\in\{0,1\}$, we will also neglect the usual discussion of vector spaces associated with fusion and splitting.)  The $F$-symbol satisfies the pentagon equation, which guarantees consistency of fusion:
\begin{equation}\label{pentagoneq}
F^{fcd}_{egl}F^{abl}_{efk}=\sum_hF^{abc}_{gfh}F^{ahd}_{egk}F^{bcd}_{khl}.
\end{equation}
The pentagon equation is illustrated in Fig.~\ref{fig:pentagon}.
We also note that $d_a$ is related to the $F$ symbol by
\begin{equation}
F^{a\bar{a}a}_{a\mathbf{1}\mathbf{1}}=\frac{\epsilon_a}{d_a}.
\end{equation}
where $\epsilon_a$ is a phase called the Frobenius-Schur indicator, that will be important later in this paper.
 
For a given set of fusion rules \eqref{fusioncat} there are in general several different solutions to the pentagon equation.   The main question we are concerned with is: how do we know when a fusion category symmetry is anomalous, given the $F$ symbol?  Of course, as described above, all solutions with $d_{a}$ non-integral are automatically anomalous, so this question is the most pressing only when the fusion rules are compatible with integral quantum dimensions. 

As reviewed above, in the group-like case, the various solutions to the pentagon equation are classified by $H^3(G,U(1))$, and the solutions giving anomalous fusion categories are those corresponding to nontrivial elements of $H^3(G,U(1))$.  In this light, in looking for a criterion for non-anomalous $F$ symbols we are asking for an analog of the $F$ symbol being a trivial cocycle.  

\begin{figure}[tb]
   \centering
   \includegraphics{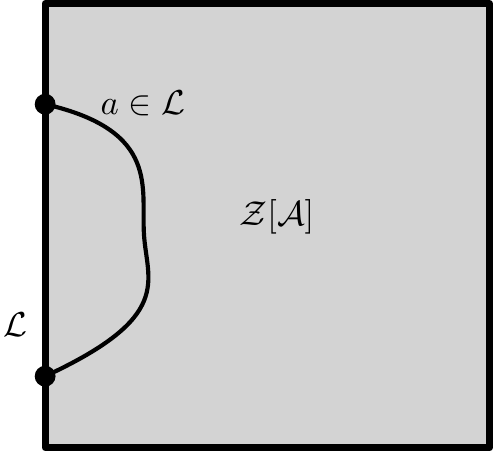} 
   \caption{We study the fusion category symmetry $\mathcal{A}$ using the Drinfeld center $\mathcal{Z}[\mathcal{A}]$ which describes a $(2+1)D$ topological order. Gapped phases of a $(1+1)D$ system with symmetry $\mathcal{A}$, which may be spontaneously broken, are in one-to-one correspondence with Lagrangian algebras $\mathcal{L}$ of $\mathcal{Z}[\mathcal{A}]$. $\mathcal{L}$ describes which anyons can be annihilated at the boundary. In this figure, $a$ is one such anyon.}
   \label{fig:bulk}
\end{figure}

A mathematical approach to this question is based on fiber functors: a fusion category symmetry is anomaly-free if and only if there exists a fiber functor, which is a tensor functor from the fusion category to $\mathrm{Vec}_{\mathbb{C}}$, the category of complex vector spaces (which has a single simple object $\mathbb{C}$) \cite{etingof2016,bhardwaj2018,thorngrenwang1}. For the purpose of just detecting when the fusion category symmetry is anomalous, this method is far too detailed and in general also too difficult.  Indeed, in physical terms, a fiber functor not only detects the absence of an anomaly, but also yields an explicit construction of a gapped, symmetric SPT phase of the given fusion category symmetry.  

In this work, we present a different approach for detecting when a fusion category symmetry $\mathcal{A}$ is anomalous. Our approach is based on certain kinds of Lagrangian algebras of the corresponding Drinfeld center $\mathcal{Z}[\mathcal{A}]$ that we call \emph{magnetic} Lagrangian algebras. For a fusion category, $\mathcal{Z}[\mathcal{A}]$ is in general a non-abelian topological order (i.e. a TQFT) (see Fig.~\ref{fig:bulk}). In the paradigm of \cite{kong2020,kaidi2022,freed2022,freed2022_2,kaidi2023} this is the symmetry TQFT, a universal gapped (2+1)D bulk that allows any possible edge physics realizing the symmetry $\mathcal{A}$.  One such boundary which always exists is gapped with domain walls labelled by the objects in $\mathcal{A}$, and hence gives a realization of the symmetry where $\mathcal{A}$ is spontaneously broken.  To detect whether or not $\mathcal{A}$ is anomalous, we ask whether the bulk topological order $\mathcal{Z}[\mathcal{A}]$ admits a gapped edge in which the categorical symmetry $\mathcal{A}$ is not spontaneously broken.\footnote{It can be technically challenging to construct the Drinfeld center of a fusion category. One way to do so is via the string-net formalism, which we use in Appendix~\ref{sstringnet}.}

We observe that $\mathcal{A}$ admits a realization in an SPT, i.e.\ is anomaly free, if and only if $\mathcal{Z}[\mathcal{A}]$ has a magnetic Lagrangian algebra.  Thus any obstruction to the existence of a magnetic Lagrangian algebra can be viewed as an anomaly for the fusion category symmetry $\mathcal{A}.$  Furthermore, there is a one-to-one correspondence between (1+1)d SPTs of the fusion category symmetry and magnetic Lagrangian algebras. This magnetic Lagrangian algebra approach can be thought of as a ``bulk" approach in contrast to the ``edge" approach of fiber functors. In particular, we will show that in many cases, there are obstructions to the existence of such a Lagrangian algebra that do not require finding all the data that specifies the Lagrangian algebra. For instance, in order for a magnetic Lagrangian algebra to exist, there must be a sufficient number of bosons in the TQFT. We will demonstrate the power of this obstruction by using it to show that certain fusion category symmetries are anomalous, without searching for Lagrangian algebras or fiber functors.

\subsection{Main result}\label{smain}
Let us now summarize our main results in more detail.  Gapped boundaries of a $(2+1)D$ TQFTs $\mathcal{C}$ are in one-to-one correspondence with Lagrangian algebras \cite{davydov2013,lan2015,cong2017}. A Lagrangian algebra $\mathcal{L}$ is a composite object formed from several different anyons:
\begin{equation}
\mathcal{L}=\oplus_an_aa,
\end{equation}
where $\{n_a\}$ are nonnegative integers and $a\in \mathcal{C}$. The formal definition of a Lagrangian algebra, which we will give in Sec.~\ref{slagrangianalgebra}, includes several additional structures and constraints. For now, we will simply use the rough physical picture that $\mathcal{L}$ defines a gapped boundary condition where anyons in $\mathcal{L}$ are simultaneously condensed, and anyons not in $\mathcal{L}$ are confined.

Many topological orders do not have any gapped boundary \cite{levin2013,kaidi2022higher}, but those that are the Drinfeld center of a fusion category $\mathcal{A}$, which we denote by $\mathcal{C}=\mathcal{Z}[\mathcal{A}]$, are guaranteed to have at least one gapped boundary. This is the canonical gapped boundary where $\mathcal{A}$ is fully spontaneously broken. The corresponding Lagrangian algebra is what we will call the \emph{electric} Lagrangian algebra $\mathcal{L}_e$. For example, consider the particular case where the fusion category $\mathcal{A}$ can be obtained from a topological order (braided fusion category) $\mathcal{A}_b$ by forgetting the braiding. In this case, $\mathcal{Z}[\mathcal{A}]=\mathcal{A}_b\boxtimes \bar{\mathcal{A}_b}$ where $\bar{\mathcal{A}_b}$ is the time-reversal of $\mathcal{A}_b$. The corresponding electric Lagrangian algebra is given by $\mathcal{L}_e=\oplus_{a\in\mathcal{A}_b}a\bar{a}$.  Relatedly, in the case where the fusion category $\mathcal{A}$ is formed out of invertible (group-like) operators, the (2+1)D bulk symmetry TQFT is a finite gauge theory and the canonical boundary condition is the Dirichlet condition.

By constrast, to show that $\mathcal{A}$ is non-anomalous, we must exhibit a gapped boundary which does not spontaneously break any symmetries in $\mathcal{A}.$ This in turn is possible if and only if in $\mathcal{Z}[\mathcal{A}]$ there exists another Lagrangian algebra, that we call a magnetic Lagrangian algebra $\mathcal{L}_m$, that intersects trivially with $\mathcal{L}_e$. Specifically, this means that the only anyon in both $\mathcal{L}_e$ and $\mathcal{L}_m$ is the vacuum anyon $\mathbf{1}$. The intuition behind this anomaly-vanishing condition is the following: because such a $\mathcal{L}_m$ does not contain any anyons in $\mathcal{L}_e$, it does not spontaneously break any of the fusion category symmetries. Therefore, $\mathcal{L}_m$ defines a symmetric, gapped $(1+1)D$ theory, which can only exist if $\mathcal{A}$ is not anomalous.  In general, it is difficult to prove that such a $\mathcal{L}_m$ exists. However, using properties of Lagrangian algebras, we can show that certain fusion categories $\mathcal{A}$ lead to obstructions to the existence of a $\mathcal{L}_m$ in $\mathcal{Z}[\mathcal{A}]$. These fusion category symmetries are therefore anomalous.

We apply this approach to Tambara-Yamagami (TY) fusion categories, which generalize the Ising fusion category (where $G=\mathbb{Z}_2$) (\ref{isingf}) to general abelian groups $G$. We focus on $G=\mathbb{Z}_N\times\mathbb{Z}_N$ with $N=2$ and $N>2,$ $N$ odd, and show that particular choices of $N$ and $F$ symbol are anomalous, in agreement with Ref.~\cite{tambara2000}.

\subsection{Example: invertible symmetry $\mathcal{A}=\mathrm{Vec}_{\mathbb{Z}_2}^{\omega}$}\label{sexamplez2}

As an example, let us consider $\mathcal{A}=\mathrm{Vec}_{\mathbb{Z}_2}^{\omega}$. This fusion category symmetry is just an invertible $\mathbb{Z}_2$ symmetry along with an $F$ symbol given by the cocycle $\omega\in H^3(\mathbb{Z}_2,U(1))=\mathbb{Z}_2$.  Of course when the $F$ symbol is non-trivial the symmetry is anomalous and cannot be realized in an invertible phase.  This is transparent from the anomaly inflow point of view where we realize the (1+1)D system as the edge of an invertible bulk phase with action described by $\omega.$  Here we instead aim to reproduce this conclusion by using the related (2+1)D Dijkgraaf-Witten theory $\mathcal{Z}[\mathcal{A}]$ which is a dynamical $\mathbb{Z}_{2}$ gauge theory with the same action $\omega$.  Specifically, we will show that simply computing the number of bosons in $\mathcal{Z}[\mathcal{A}]$ detects an anomaly. 

This fusion category $\mathcal{A}$ has only two objects $\{1,g\}$, with the fusion rules
\begin{equation}
a\times\mathbf{1}=a\qquad g\times g=\mathbf{1},
\end{equation}
for all $a\in\mathcal{A}$.  Additionally, we must specify the $F$ symbol. Explicitly, this is a function from $\mathbb{Z}_2\times\mathbb{Z}_2\times\mathbb{Z}_2\to U(1)$ satisfying the cocycle condition/pentagon equation \eqref{pentagoneq} . Let us denote the trivial cocycle by $\omega_0$ and the nontrivial cocycle by $\omega_1$. In particular, $\omega_1(g,g,g)=-1$, which pictorially means that
\begin{equation}
\includegraphics{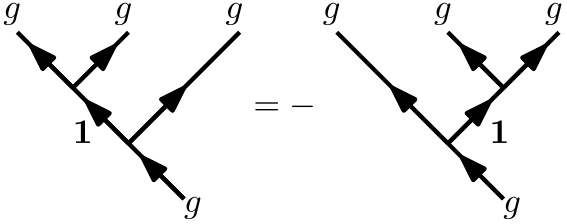}
\end{equation}
$\mathcal{Z}[\mathrm{Vec}_{\mathbb{Z}_2}^{\omega}]$ is an abelian $\mathbb{Z}_2$ gauge theory, with four anyons $\{\mathbf{1},e,m,em\}$. Here $e$ denotes the electric Wilson line, and $m$ its magnetic dual (the $\mathbb{Z}_2$ flux).  The two choices of cocycle $\omega$ give are distinguished by the topological spin of $m$ (and therefore also the topological spin of $em$): $\omega_{0}$ yields untwisted $\mathbb{Z}_{2}$ gauge theory with bosonic $m$, while $\omega_{1}$ yields twisted $\mathbb{Z}_{2}$ gauge theory with semionic $m$ (topological spin $i$). 

To exhibit the anomaly, we must now check whether these topological field theories admit gapped boundaries.  Since these theories are abelian, we do not need all of the structure of Lagrangian algebras and can instead use Lagrangian subgroups \cite{kapustin2011,levin2013}. A Lagrangian subgroup $\tilde{\mathcal{L}}$ is a subgroup of the abelian anyons satisfying the following two properties:
\begin{enumerate}
\item{The anyons in $\tilde{\mathcal{L}}$ have trivial mutual statistics: $\theta_{ab}=1$ for all $a,b\in\tilde{\mathcal{L}}$. In particular, for bosonic topological orders, $\theta_a=1$ for all $a\in \tilde{\mathcal{L}}$, where $\theta_a$ is the topological spin of $a$. This ensures that the anyons in $\tilde{\mathcal{L}}$ can be simultaneously condensed.}
\item{Any anyon not in $\tilde{\mathcal{L}}$ has nontrivial mutual statistics with at least one anyon in $\tilde{\mathcal{L}}$. This ensures that all anyons not in $\tilde{\mathcal{L}}$ are confined.}
\end{enumerate}

In addition, the number of anyons in $\tilde{\mathcal{L}}$, $|\mathcal{L}|$, satisfies 
\begin{equation}
|\mathcal{L}|=\mathcal{D}=\sqrt{\sum_{a\in\mathcal{C}}d_a^2}=\sqrt{|\mathcal{C}|},
\end{equation}
where $|\mathcal{C}|$ is the number of anyons in $\mathcal{C}$. Here, we used the fact that $d_a=1$ for all $a\in\mathcal{C}$ if $\mathcal{C}$ is  abelian. 

With the trivial cocycle, the resulting $\mathbb{Z}_2$ gauge theory, has two Lagrangian subgroups:
\begin{equation}
\tilde{\mathcal{L}}_e=\{1,e\}\qquad \tilde{\mathcal{L}}_m=\{1,m\}.
\end{equation}
The first of these is the expected electric Lagrangian subgroup, which condenses to give a gapped edge with the $\mathbb{Z}_{2}$ symmetry spontaneously broken.  The second is the magnetic Lagrangian subgroup which condenses to give a gapped edge without spontaneous symmetry breaking.  On the other hand, twisted $\mathbb{Z}_2$ gauge theory has only one Lagrangian subgroup: 
\begin{equation}
\tilde{\mathcal{L}}_e=\{1,e\}.
\end{equation}
In particular, because the other two anyons in twisted $\mathbb{Z}_2$ gauge theory have topological spin $i$ and $-i$, they cannot be condensed. Therefore, there is no magnetic Lagrangian subgroup in twisted $\mathbb{Z}_2$ gauge theory.

From the analysis above, we recover the fact that $\mathrm{Vec}_{\mathbb{Z}_2}^{\omega_0}$ is anomaly free, while $\mathrm{Vec}_{\mathbb{Z}_2}^{\omega_1}$ is anomalous. However, even without explicitly finding all Lagrangian subgroups, we can already easily detect that twisted $\mathbb{Z}_{2}$ gauge theory does not have a $\tilde{\mathcal{L}}_m$ by simply examining anyon spins.  Indeed, for a $\tilde{\mathcal{L}}_m$ to exist, there to be at least two anyons not including the gauge charge $e$ which are bosons. By inspecting the topological spins of anyons in twisted $\mathbb{Z}_2$ gauge theory, we see that there are not enough bosons. Disregarding $e$, there is only a single bosonic anyon in the theory, which is $\mathbf{1}$. Therefore, the theory cannot possibly have a $\tilde{\mathcal{L}}_m$.

\section{Review of $(2+1)D$ topological orders}\label{sprelim}

Much of the analysis above carries over to more general fusion categories $\mathcal{A}$ that describe non-invertible symmetries, whose Drinfeld centers $\mathcal{Z}[\mathcal{A}]$ are non-abelian topological orders. In this section, we briefly review relevant background before applying these ideas to anomalies of fusion category symmetries.   For more details, see \cite{kitaev2006,etingof2016,barkeshli2019}.

\subsection{Braided fusion categories}\label{scenter}

A $(2+1)D$ topological order is defined mathematically as a unitary modular tensor category $\mathcal{C}$ \cite{kitaev2006}, which is a category with fusion and braiding described by $F$ and $R$ symbols respectively, defined in a consistent way. For the application to anomalies, we are interested in particular in Drinfeld centers of fusion categories, but in this section we will consider general $\mathcal{C}.$

We now proceed to briefly review some properties of general topological orders, that are not necessarily of the form $\mathcal{Z}[\mathcal{A}]$ for some fusion category $\mathcal{A}$. A topological order $\mathcal{C}$ is described by a set of simple objects $\{a\}$ that we call anyons. Note that when $\mathcal{C}=\mathcal{Z}[\mathcal{A}]$, this set is different from but related to the simple objects in $\mathcal{A}$. The quantum dimensions of the anyons define a total quantum dimension $\mathcal{D}$ of $\mathcal{C}$:
\begin{equation}
\mathcal{D}=\sqrt{\sum_{a\in \mathcal{C}}d_a^2}.
\end{equation}

$\mathcal{C}$ is equipped with both fusion and braiding, described by $\tilde{F}^{abc}_{def}$ and $R^{ab}_c$ respectively. Here we use the notation $\tilde{F}^{abc}_{def}$ to differentiate it from the $F$ symbol of the fusion category $F^{abc}_{def}$. Braiding, given by the $R$ symbol, describes the following process:
\begin{equation}
\includegraphics{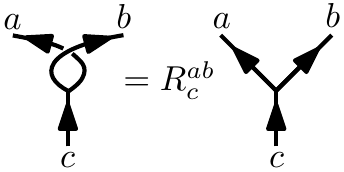}.
\end{equation}

In order for fusion and braiding to be consistent, $\tilde{F}^{abc}_{def}$ and $R^{ab}_c$ must satisfy the hexagon equations, which we will omit because they are not necessary for this work. Braiding defines the topological spin of each anyon:
\begin{equation}\label{selfstat}
\theta_a=\theta_{\bar{a}}=\sum_c\frac{d_c}{d_a}R^{aa}_c, 
\end{equation}
and it can be shown that mutual braiding and topological spin are related in the following way\cite{kitaev2006}:
\begin{equation}\label{braidstat}
R^{ab}_cR^{ba}_c=\frac{\theta_c}{\theta_a\theta_b}.
\end{equation}

The quantum dimensions and topological spins of the anyons define the topological $S$ and $T$ matrices. The $S$ matrix is defined as
\begin{equation}
S_{ab}=\frac{1}{\mathcal{D}}\sum_cN_{\bar{a}b}^c\frac{\theta_c}{\theta_a\theta_b}d_c.
\end{equation}
The $S$ matrix is useful for computing the fusion coefficients from the Verlinde formula:
\begin{equation}\label{verlinde}
N_{ab}^c=\sum_{x\in\mathcal{C}}\frac{S_{ax}S_{bx}S_{cx}^*}{S_{0x}}.
\end{equation}
The $T$ matrix is given by
\begin{equation}
T_{ab}=\theta_a\delta_{ab}.
\end{equation}
For a topological order of the form $\mathcal{Z}[\mathcal{A}]$, the $S$ and $T$ matrices are always unitary and thus nondegenerate. This means that all the anyons are detectable via braiding: for each anyon $a\in\mathcal{C}$ other than $\mathbf{1}$, there exists another anyon $b\in\mathcal{C}$ that braids nontrivially with $a$.\footnote{In particular, this means that we are studying bosonic topological orders which do not have transparent fermions.} 

\subsection{Lagrangian algebras}\label{slagrangianalgebra}

 In this section, we give a complete definition of Lagrangian algebras, which describe gapped boundaries of general topological orders \cite{davydov2013,kong2014,cong2017}. Our treatment follows \cite{cheng2020}.

A Lagrangian algebra $\mathcal{L}=\oplus_an_aa$ describes which anyons can be annihilated at the boundary of a topological order with local operators. The multiplicity $n_a$, which is a nonnegative integer, specifies the number of inequivalent ways $a$ can be annihilated at the boundary. It follows that $n_a$ is the dimension of ``boundary condensation space" $V^a$, which is the vector space $\{|a;\mu\rangle\}$ for local operators that annihilate $a$. $\mathcal{L}$ must include $\mathbf{1}$, with $n_{\mathbf{1}}=1$. An additional piece of data describing $\mathcal{L}$ is an algebra morphism $\mathcal{L}\times\mathcal{L}\to\mathcal{L}$ given by the $M$ symbol, which describes the following physical process:
\begin{equation}
\includegraphics{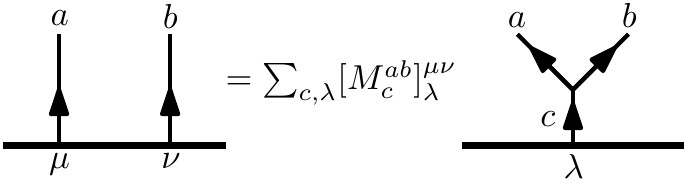}
\end{equation}
where the bottom bold line is the boundary. Here, $\mu,\nu,$ and $\lambda$ are states in the boundary condensation spaces of $a,b,$ and $c$ respectively.

\begin{figure}[tb]
   \centering
   \includegraphics{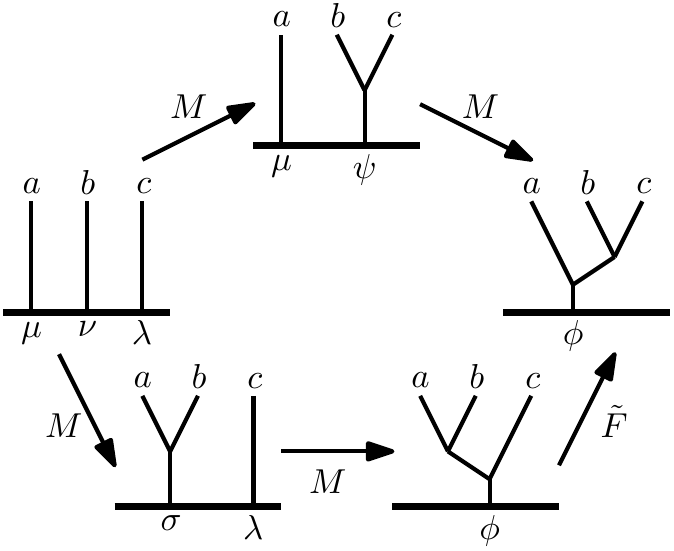} 
   \caption{The boundary pentagon equation is a consistency equation ensuring the two paths to the same configuration given match. We have again omitted the arrows on the anyon lines for clarity of the figure.}
   \label{fig:pentagonbdry}
\end{figure}

Unlike the $\tilde{F}$ and $R$ symbols, the $M$ symbol is not unitary; it does not preserve the dimensions of the vector spaces on the left and right hand side. Instead, as shown in \cite{cong2017},
\begin{equation}\label{ineq}
n_an_b\leq\sum_cN_{ab}^cn_c.
\end{equation}
Consistency of the $M$ symbol together with the $\tilde{F}$ symbol gives the following variant of the pentagon equation:
\begin{equation}\label{mfuse}
\sum_{e,\sigma}\left[M^{ab}_c\right]^{\mu\nu}_{\sigma}\left[M^{ec}_d\right]^{\sigma\lambda}_{\delta}\tilde{F}^{abc}_{def}=\sum_{\psi}\left[M^{af}_d\right]^{\mu\psi}_{\delta}\left[M^{bc}_f\right]^{\nu\lambda}_{\psi},
\end{equation}
which is illustrated pictorially in Figure \ref{fig:pentagonbdry}.

There are additional constraints on the $M$ symbol related to braiding. Note that an anyon in $\mathcal{L}$ disappears when it is annihilated at the boundary, so it should not matter in what order the anyons are annihilated at the boundary. This means that
\begin{equation}\label{mbraid}
\left[M^{ba}_c\right]^{\nu\mu}_{\lambda}R^{ab}_c=\left[M^{ab}_c\right]^{\mu\nu}_{\lambda},
\end{equation}
which pictorially means
\begin{equation}
\includegraphics{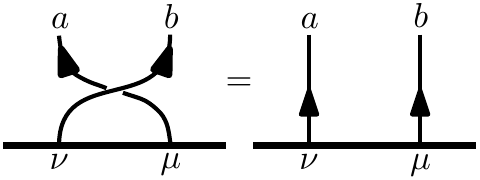}.
\end{equation}

From (\ref{mbraid}), it is easy to see that
\begin{equation}\label{braidl}
R^{ab}_cR^{ba}_c=1,
\end{equation}
if $a,b,$ and $c$ are all in $\mathcal{L}$. Furthermore, for bosonic topological orders like the ones considered in this work, all the anyons in $\mathcal{L}$ are bosons.  Finally, we also have the condition
\begin{equation}\label{dim}
\sum n_ad_a=\mathcal{D}.
\end{equation}

With an eye towards computing anomalies, we note that searching for an $\mathcal{L}$ satisfying \eqref{ineq} and \eqref{dim} composed only of bosonic anyons with trivial braiding within channels in $\mathcal{L}$ \eqref{braidl} is often straightforward. Hence failure of these conditions can be used to obstruct the existence of an $\mathcal{L}_m$ and prove that a fusion category is anomalous.  This will be our main strategy below.\footnote{These conditions are not sufficient for $\mathcal{L}$ to define a Lagrangian algebra as one must also construct a consistent $M$ symbol. }  

Let us give some examples of Lagrangian algebras. As mentioned in Sec.~\ref{smain}, for a topological order of the form $\mathcal{Z}[\mathcal{A}]=\mathcal{A}_b\boxtimes \bar{\mathcal{A}_b}$, the Lagrangian algebra corresponding to the boundary where $\mathcal{\mathcal{A}}$ is fully spontaneously broken is given by $\mathcal{L}_e=\oplus_{a\in\mathcal{A}}a\bar{a}$. For example, for the doubled Ising topological order $\mathcal{C}=\{\mathbf{1},\psi,\bar{\psi},\sigma,\bar{\sigma},\psi\bar{\sigma},\sigma\bar{\psi},\psi\bar{\psi},\sigma\bar{\sigma}\}$, we have
\begin{equation}
\mathcal{L}_e=\mathbf{1}+\psi\bar{\psi}+\sigma\bar{\sigma}.
\end{equation}
Similarly, for the doubled Fibonacci topological order $\mathcal{C}=\{\mathbf{1},\tau,\bar{\tau},\tau\bar{\tau}\}$, we have
\begin{equation}
\mathcal{L}_e=\mathbf{1}+\tau\bar{\tau}.
\end{equation}
Note that both the doubled Ising topological order and the doubled Fibonacci topological topological order only have a single Lagrangian algebra, which is the $\mathcal{L}_e$. Because there does not exist a $\mathcal{L}_m$ in either theory, the Ising fusion category and the Fibonacci fusion category are both anomalous. This agrees with the simple analysis using fusion rules discussed around \eqref{qdanom}, since $\sigma$ and $\tau$ both have non-integral quantum dimension.

Another class of non-abelian topological orders are those of the form $\mathcal{Z}[\mathrm{Rep}_G]$ where $G$ is a non-abelian group. Here, the electric Lagrangian algebra is given by $\mathcal{L}_e=\oplus_{a\in\mathrm{Rep}_G}d_aa$ where $d_a$ is the dimension of the irreducible representation $a$ of $G$ (and also the quantum dimension of the anyon $a$).\footnote{Recall that anyons of $\mathcal{Z}[\mathrm{Rep}_G]$ are labeled by $(c,\rho_u)$ where $c$ is a conjugacy class of $G$, $u$ is an element of $c$, and $\rho_u$ is an irreducible representation of the centralizer of $u$ in $G$. $G$ always has the conjugacy class $\{1\}$, and the centralizer of $1$ is the whole group. By irreducible representations $a$ here, we mean anyons of the kind $(\{1\},\rho_1).$} This is the familiar Dirichlet boundary condition for the gauge theory. More generally, for any topological order of the form $\mathcal{Z}[\mathcal{A}]$, there always exists an $\mathcal{L}_e$. In the string-net formalism \cite{levin2005}, the set of $a\in\mathcal{C}$ in $\mathcal{L}_e$ are those that contain the vacuum string, because these are the anyons that can be absorbed into the ``smooth" boundary (fully spontaneously broken) of the string-net model. For an example of finding $\mathcal{L}_e$ from the string-net formalism, see Appendix~\ref{sstringnet}. 
\section{Warmup: abelian topological orders}\label{sabelian}
In this section, we derive two well-known results using magnetic Lagrangian subgroups. First, we will show that $\mathcal{Z}[\mathrm{Vec}_{G}^{\omega}]$ is anomalous whenever $G$ is a finite abelian group and $\omega$ is a type-I or type-II cocycle, as a generalization of the analysis in Sec.~\ref{sexamplez2}. (We will soon explain what we mean by ``type-I" and ``type-II".) We will consider in particular $G=\mathbb{Z}_{N_1}\times\mathbb{Z}_{N_2}$, which is the simplest group that can have both a nontrivial type-I cocycle and a type-II cocycle, but the analysis straightforwardly generalizes to general finite abelian $G$.  Second, we show that there is a $\mathbb{Z}_N$ classification of $(1+1)D$ SPTs when $G=\mathbb{Z}_N\times\mathbb{Z}_N$ and $\omega$ is trivial.  Again this analysis also extends to general $G$.

\subsection{Anomalies from abelian topological orders}\label{stype1}
Let us consider the topological order $\mathcal{Z}[\mathrm{Vec}_{\mathbb{Z}_{N_1}\times\mathbb{Z}_{N_2}}^{\omega}]$ where $\omega\in H^3(\mathbb{Z}_{N_1}\times\mathbb{Z}_{N_2},U(1))$. These kinds of 3-cocycles are built out of two types, labeled type-I and type-II \cite{propitius1995,wang2015,jwang2015}. A type-I cocycle involves a single cyclic group and modifies the topological spin of the gauge fluxes of that cyclic group. On the other hand, a type-II cocycle involves two cyclic groups and modifies the braiding of the gauge fluxes of the two cyclic groups. It therefore describes a mixed anomaly between the two cyclic groups. 

$\mathcal{Z}[\mathrm{Vec}_{\mathbb{Z}_{N_1}\times\mathbb{Z}_{N_2}}^{\omega}]$ has $|N_1N_2|^2$ anyons, generated by $\{e_1,m_1,e_2,m_2\}$. A generic anyon in the theory is labeled by $e_1^{p_1}m_1^{q_1}e_2^{p_2}m_2^{q_2}$, where $p_i,q_i\in[0,N_i-1]$. The $\tilde{\mathcal{L}}_e$ (recall that we use $\tilde{\mathcal{L}}$ to denote Lagrangian subgroup and $\mathcal{L}$ to denote Lagrangian algebra) is generated by $\{e_1,e_2\}$, i.e.
\begin{equation}
\tilde{\mathcal{L}}_e=\{e_1^{p_1}e_2^{p_2}\},
\end{equation}
for $p_i\in[0,N_i-1]$.

Before we describe a useful necessary condition for there to be a $\tilde{\mathcal{L}}_m$, it is convenient to define an equivalence relation where we ``mod out" by gauge charges. We define:
\begin{equation}\label{equivm}
e_1^{p_1}m_1^{q_1}e_2^{p_2}m_2^{q_2}\sim m_1^{q_1}m_2^{q_2}.
\end{equation}
Using this equivalence relation, we can state a necessary condition for there being a $\tilde{\mathcal{L}}_m$ as the following: there needs to be a bosonic anyon within each equivalence class of anyons under (\ref{equivm}). To see why this is the case, suppose that instead, we construct a Lagrangian subgroup with two elements from the same equivalence class under (\ref{equivm}):
\begin{equation}
\{e_1^{p_1}m_1^{q_1}e_2^{p_2}m_2^{q_2},e_1^{p_1'}m_1^{q_1}e_2^{p_2'}m_2^{q_2}\}\in\tilde{\mathcal{L}}
\end{equation}

Then by fusing the first anyon with the antiparticle of the second, we see that they generate $e_1^{p_1-p_1'}e_2^{p_2-p_2'}$, which must also be in $\tilde{\mathcal{L}}$. But this is a pure charge, so it means that $\tilde{\mathcal{L}}$ intersects nontrivially with $\tilde{\mathcal{L}}_e$, so it fails to be a magnetic Lagrangian subgroup. In order for there to exist a magnetic Lagrangian subgroup with $|N_1N_2|$ anyons, we need at least one element of each equivalence class under (\ref{equivm}) to be a boson.

We will now show that if $\omega$ is a nontrivial 3-cocycle, then we cannot obtain a boson in each equivalence class. To do so, we will first review how $\omega$ specifies the topological spins of the gauge fluxes.

The topological spin of $m_i$ is determined by $\omega$ \cite{wang2015}:
\begin{equation}\label{nspin}
\theta_{m_i}^{N_i}=\prod_{n=0}^{N_i-1}\omega(g_i,g_i^n,g_i)= e^{\frac{2\pi ik_i}{N_i}},
\end{equation}
where $k_i\in[0,N_i-1]$ distinguishes the different equivalence classes of type-I cocycles and $k\neq 0$ if $\omega$ is nontrivial. From (\ref{nspin}) and (\ref{braidstat}), we see that
\begin{equation}\label{mstattype1}
\theta_{m_i^q}=e^{\frac{2\pi iq^2k_i}{N_i^2}}.
\end{equation}
The mutual statistics of $m_1$ and $m_2$ is also given by $\omega$:
\begin{equation}\label{rm1m2}
\left(R^{m_1,m_2}R^{m_2,m_1}\right)^{N^{12}}=e^{\frac{2\pi i k_{12}}{N_{12}}},
\end{equation}
where $N^{12}=\mathrm{lcm}(N_1,N_2)$ and $N_{12}=\mathrm{gcd}(N_1,N_2)$. $k_{12}\in[0,N_{12}-1]$ distinguishes the different equivalence class of type-II cocycles. In particular, $k_{12}\neq 0$ if there is a nontrivial mixed anomaly between the two cyclic groups. Note that we omit the explicit expression for $e^{\frac{2\pi ik_{12}}{N_{12}}}$ in terms of $\omega$ here; it can be found in \cite{wang2015}. 

Using (\ref{braidstat}) together with (\ref{rm1m2}), we have
\begin{equation}
\theta_{m_1^{q_1}m_2^{q_2}}=e^{\frac{2\pi ik_{12}q_1q_2}{N_1N_2}}e^{\frac{2\pi ik_1q_1^2}{N_1^2}}e^{\frac{2\pi ik_2q_2^2}{N_2^2}}.
\end{equation}

Now that we have all the information about the topological spins of the gauge fluxes, we need to check if there exists an anyon in each equivalence class under (\ref{equivm}) that is bosonic. First, let us check that there exists an anyon in each equivalence class labeled by $m_i^{q_i}$ with bosonic topological spin. In particular, there must be a bosonic anyon with $q_i=1$, so there must be a solution to
\begin{equation}
\theta_{e_1^{p_1}e_2^{p_2}m_i}=\theta_{e_i^{p_i}m_i}=e^{\frac{2\pi ik_i}{N_i^2}}e^{\frac{2\pi ip_i}{N_i}}=1.
\end{equation}

Clearly, there is no solution to the above equation if $0<k_i<N_i$, so if $k_1\neq 0$ or $k_2\neq 0$, then there cannot exist a magnetic Lagrangian subgroup.

Now suppose that $k_1=k_2=0$. We must also check that there exists an anyon in the equivalence class labeled by $m_1^{q_1}m_2^{q_2}$ that has bosonic topological spin. In particular, there must be a bosonic anyon with $q_1=q_2=1$. We have
\begin{equation}
\theta_{e_1^{p_1}m_1e_2^{p_2}m_2}=e^{\frac{2\pi ik_{12}}{N_1N_2}}e^{\frac{2\pi ip_1}{N_1}}e^{\frac{2\pi i p_2}{N_2}}=1.
\end{equation}

If $0<k_{12}<N_{12}$, then the above equation cannot hold for any $p_1$ and $p_2$, so there cannot exist a magnetic Lagrangian subgroup. 

In summary, any twisted $\mathbb{Z}_{N_1}\times\mathbb{Z}_{N_2}$ gauge theory defined by a nontrivial $\omega\in H^3(\mathbb{Z}_{N_1}\times\mathbb{Z}_{N_2},U(1))$ cannot have a magnetic Lagrangian subgroup, recovering the known fact that $\mathrm{Vec}_{\mathbb{Z}_{N_1}\times\mathbb{Z}_{N_2}}^\omega$ is anomalous when the $F$ symbol (given by $\omega$) is a nontrivial element of $H^3(\mathbb{Z}_{N_1}\times\mathbb{Z}_{N_2},U(1))$.

\subsection{$(1+1)D$ SPTs with $\mathbb{Z}_N\times\mathbb{Z}_N$ symmetry}\label{sspt1d}
We will now study the fusion category $\mathrm{Vec}_{\mathbb{Z}_N\times\mathbb{Z}_N}^{\omega}$ where $\omega$ is a trivial 3-cocycle. We will show that there are $N$ distinct $\tilde{\mathcal{L}}_m$, so there are $N$ SPTs with $\mathbb{Z}_N\times\mathbb{Z}_N$ symmetry in $(1+1)D$. 

$\mathcal{Z}[\mathrm{Vec}_{\mathbb{Z}_N\times\mathbb{Z}_N}^{1}]$ has $N^4$ anyons, generated by $\{e_1,m_1,e_2,m_2\}$. The $\mathcal{L}_m$ are given by
\begin{equation}\label{lmp}
\tilde{\mathcal{L}}_{m}^{(p)}=\left\{\left(m_1e_2^p\right)^s\left(m_2e_1^{-p}\right)^t\right\}.
\end{equation}
where $s,t\in[0,N-1]$. Let us check that $\tilde{\mathcal{L}}_m^{(p)}$ is a valid Lagrangian subgroup. First, because $e_1$ braids trivially with $m_2$ and $e_2$ braids trivially with $m_1$, the two generators of $\tilde{\mathcal{L}}_m^{(p)}$ are bosonic: $\theta_{m_1e_2^p}=\theta_{m_2e_1^{-p}}=1$. Second, $m_1e_2^p$ and $m_2e_1^{-p}$ have trivial mutual statistics. 
 Therefore, all the anyons generated by $m_1e_2^p$ and $m_2e_1^{-p}$, which are all the anyons in $\tilde{\mathcal{L}}_m^{(p)}$, are bosonic. It is also clear that $|\tilde{\mathcal{L}}_m^{(p)}|=N^2=\mathcal{D}$, where $\mathcal{D}$ is the total quantum dimension of $\mathcal{Z}[\mathrm{Vec}_{\mathbb{Z}_N\times\mathbb{Z}_N}^{1}]$. Finally, there are no pure gauge charges in $\tilde{\mathcal{L}}_m^{(p)}$: there is a single element from each equivalence class under fusion with charges.

Any other Lagrangian subgroup would overlap nontrivially with $\tilde{\mathcal{L}}_e$. This is because the only other way to get a Lagrangian subgroup that contains a single element from each equivalence class under fusion with charges (and hence does not contain any pure charges) is to instead generate it by $m_1e_2^{p_2}$ and $m_2e_1^{p_1}$ where $p_2\neq -p_1$ mod $N$. However, then $m_1e_2^{p_2}m_2e_1^{p_1}$ is not a boson, so these anyons cannot generate a Lagrangian subgroup. Therefore, the only $\tilde{\mathcal{L}}_m$ are those given in (\ref{lmp}). Since $p$ can take $N$ values, there are $N$ distinct $\tilde{\mathcal{L}}_m$ and hence $N$ $(1+1)D$ $\mathbb{Z}_N\times\mathbb{Z}_N$ SPTs.

For concreteness, consider the particular case where $N=2$. The Lagrangian subgroups of this topological order are given by
\begin{align}
\begin{split}\label{lz2z2}
\tilde{\mathcal{L}}_e&=\{\mathbf{1},e_1,e_2,e_1e_2\}\\
\tilde{\mathcal{L}}_a&=\{\mathbf{1},e_1,m_2,e_1m_2\}\\
\tilde{\mathcal{L}}_b&=\{\mathbf{1},m_1,e_2,m_1e_2\}\\
\tilde{\mathcal{L}}_c&=\{\mathbf{1},e_1e_2,m_1m_2,e_1e_2m_1m_2\}\\
\tilde{\mathcal{L}}_{m}^{(0)}&=\{\mathbf{1},m_1,m_2,m_1m_2\}\\
\tilde{\mathcal{L}}_{m}^{(1)}&=\{\mathbf{1},e_1m_2,e_2m_1,e_1m_2e_2m_1\}.
\end{split}
\end{align}

$\tilde{\mathcal{L}}_e$ corresponds to a $(1+1)D$ gapped boundary with the full $\mathbb{Z}_2\times\mathbb{Z}_2$ symmetry spontaneously broken, while $\tilde{\mathcal{L}}_a,\tilde{\mathcal{L}}_b,$ and $\tilde{\mathcal{L}}_c$ correspond to boundaries that spontaneously break different $\mathbb{Z}_2$ subgroups of $\mathbb{Z}_2\times\mathbb{Z}_2$. Notice that $\tilde{\mathcal{L}}_{m}^{(0)}$ and $\mathcal{L}_{m}^{(1)}$ do not contain any pure charges. Therefore, $\tilde{\mathcal{L}}_{m}^{(0)}$ and $\tilde{\mathcal{L}}_{m}^{(1)}$ correspond to gapped boundaries that do not spontaneously break any of the $\mathbb{Z}_2\times\mathbb{Z}_2$ symmetries. These give the two $\mathbb{Z}_2\times\mathbb{Z}_2$ SPTs in $(1+1)D$.

\section{Anomalies of Tambara-Yamagami fusion categories}\label{sty}
In this section, we will analyze anomalies of Tambara-Yamagami (TY) fusion categories using magnetic Lagrangian algebras. We denote these fusion categories by $\mathrm{TY}_{G}^{\chi,\epsilon}$, where $G$ is an abelian group and $\chi$ and $\epsilon$ are defined below and specify the $F$ symbol. We will focus on $G=\mathbb{Z}_N\times\mathbb{Z}_N$, because $\mathrm{TY}_{\mathbb{Z}_N}^{\chi,\epsilon}$ is always anomalous even without specifying the $F$ symbol (see Sec.~\ref{sinvariantgroup}). After reviewing $\mathrm{TY}_G^{\chi,\epsilon}$, we will use the method of gauging a 0-form duality symmetry to obtain some information and intuition about $\mathcal{Z}[\mathrm{TY}_{\mathbb{Z}_N\times\mathbb{Z}_N}^{\chi,\epsilon}]$. We compute the topological spins of the anyons in $\mathcal{Z}[\mathrm{TY}_{G}^{\chi,\epsilon}]$ using the string-net formalism \cite{levin2005,lin2021} in Appendix~\ref{sstringnet}. We will show that $\mathcal{Z}[\mathrm{TY}_{\mathbb{Z}_N\times\mathbb{Z}_N}^{\chi,\epsilon}]$ with particular choices of $N,\chi$ and $\epsilon$ are anomalous. Our results agree with those of \cite{tambara2000}, which approached the problem using fiber functors, where they overlap.

\subsection{Tambara-Yamagami fusion categories}\label{styfusion}
The simple objects of $\mathrm{TY}_{G}^{\chi,\epsilon}$ are given by elements of an abelian group $G$, along with a duality object $\sigma$ \cite{tambara1998}. The objects have the following fusion rules:
\begin{align}
\begin{split}\label{fusion}
g\times \sigma&=\sigma\times g=\sigma,\\ \sigma\times\sigma&=\sum_{g\in G}g.
\end{split}
\end{align} 

The only nontrivial $F$ symbols are $F^{g\sigma h}_{\sigma\sigma\sigma},F^{\sigma g\sigma}_{h\sigma\sigma},$ and $F^{\sigma\sigma\sigma}_{\sigma gh}$. These $F$ symbols are given by
\begin{align}
\begin{split}\label{fsymbol}
F^{g\sigma h}_{\sigma\sigma\sigma}&=F^{\sigma g\sigma}_{h\sigma\sigma}=\chi(g,h)\\
F^{\sigma\sigma\sigma}_{\sigma gh}&=\frac{\epsilon}{\sqrt{|G|}}\chi(g,h)^{-1},
\end{split}
\end{align}
where $\epsilon=\pm1$ is the Frobenius-Schur indicator, which we will discuss more in the next section. In order for the $F$ symbol to satisfy the pentagon equation (\ref{pentagoneq}), $\chi(g,h)$ must be a function from $G\times G\to U(1)$ with the following properties:
\begin{align}
\begin{split}\label{bicharacter}
\chi(g,h)&=\chi(h,g)\\
\chi(gh,k)&=\chi(g,k)\chi(h,k)\\
\chi(g,hk)&=\chi(g,h)\chi(g,k).
\end{split}
\end{align}

In other words, $\chi(g,h)$ is a symmetric bicharacter. Equations \eqref{fusion}-\eqref{bicharacter} completely define the fusion category $\mathrm{TY}_{G}^{\chi,\epsilon}$. 

For $G=\mathbb{Z}_N\times\mathbb{Z}_N$, (\ref{bicharacter}) means that $\chi(g,h)$ is completely specified by the following values\footnote{Note that we are not allowed to choose any $x,y,z$, because for example if we choose $x=y=z=1$, the $F$ symbol is not unitary. A closer inspection on the constraints due to unitarity of the $F$ symbol shows that we can only choose $x,y,z$ satisfying
\begin{align}
\begin{split}\label{constraint}
\frac{1}{N}\sum_{h_1,h_2=0}^{N-1}&x^{h_1(g_1-k_1)}y^{h_2(g_2-k_2)}z^{h_1(g_2-k_2)+h_2(g_1-k_1)}\\
&=\delta_{g_1,k_1}\delta_{g_2,k_2}.
\end{split}
\end{align}

Clearly, the diagonal bicharacter ($x=y=e^{\frac{2\pi i}{N}},z=1$) and the off-diagonal bicharacter ($x=y=1,z=e^{\frac{2\pi i}{N}}$) are solutions to (\ref{constraint}).}:
\begin{align}
\begin{split}\label{bicharacterxyz}
x&=\chi((1,0),(1,0))\\
y&=\chi((0,1),(0,1))\\
z&=\chi((1,0),(0,10)),
\end{split}
\end{align}
where $(1,0)$ and $(0,1)$ are the two generators of $\mathbb{Z}_N\times\mathbb{Z}_N$. We will focus on two particular bicharacters in this work, although our methods apply to general bicharacters. We define the \emph{diagonal} bicharacter as
 \begin{equation}\label{diagxyz}
x=y=e^{\frac{2\pi i}{N}}\qquad z=1,
\end{equation}
and we define the \emph{off-diagonal} bicharacter as
\begin{equation}\label{offdiagxyz}
x=y=1\qquad z=e^{\frac{2\pi i}{N}}.
\end{equation}

Let us also comment on the physical realization of this class of fusion categories.  The duality object $\sigma$ naturally arises in theories which are invariant under gauging the $G$ symmetry.  Notice that this is possible in (1+1)D since the twisted sectors produced by gauging give rise to new 0-form $G$ symmetry operators.  Particular examples realizing this construction are the critical Ising model (for $G=\mathbb{Z}_{2}$) and the parafermion coset model $SU(2)_{N}/U(1)$ (for $G=\mathbb{Z}_{N}$).  Other examples include the compact boson at special values of the radius.  There are also gapped theories that are invariant under gauging the $G$ symmetry (see e.g.\ \cite{Huang:2021zvu}). These theories are the focus of this work.

\subsection{$\mathcal{Z}[\mathrm{TY}_{G}^{\chi,\epsilon}]$ from gauging duality symmetry}\label{sset}

We will now derive the properties of $\mathcal{Z}[\mathrm{TY}_{G}^{\chi,\epsilon}]$ using the method of gauging a duality symmetry in (untwisted) $G$ gauge theory, which is the abelian topological order $\mathcal{Z}[\mathrm{Vec}_{G}^1]$\cite{barkeshli2019,kaidi2022}. We can also obtain $\mathcal{Z}[\mathrm{TY}_{G}^{\chi,\epsilon}]$ directly from the input fusion category using the string-net formalism \cite{levin2005,lin2021} as shown in Appendix~\ref{sstringnet}. Here we will focus on $G=\mathbb{Z}_N\times\mathbb{Z}_N$ for $N=2$ and $N>2,N$ odd to simplify the calculation of the topological spins and obtain the results in Appendix~\ref{ssum}, but one can generalize the analysis to any finite abelian group. 

\subsubsection{Duality-enriched $G$ gauge theory}\label{sdualityenriched}
Let us consider $(2+1)D$ $\mathbb{Z}_N\times\mathbb{Z}_N$ gauge theory enriched with a duality symmetry, i.e., a global 0-form $\mathbb{Z}_2$ symmetry. This symmetry gives an order two permutation of the anyons that maintains the topological spins of the anyons (and hence also the braiding statistics) \cite{barkeshli2019}. Following \cite{barkeshli2019}, we will use the short-hand notation $^{\mathbf{g}}a$ to indicate the anyon obtained by acting on $a$ with a 0-form symmetry $\mathbf{g}$. In equation form, the above statement means that
\begin{equation}\label{permutetheta}
\theta_{a}=\theta_{^{\mathbf{g}}a}.
\end{equation}
We call a $\mathbb{Z}_2$ symmetry whose only action is to permute anyons a \emph{duality} symmetry.\footnote{\label{foot1} In general, one must also keep track of the symmetry fractionalization characterized by the degree two cohomology class from the 0-form symmetry ($\mathbb{Z}_{2}$) to the 1-form symmetry/abelian anyons ($\mathbb{Z}_{N}\times \mathbb{Z}_{N}\times \mathbb{Z}_{N}\times \mathbb{Z}_{N}$).  For examples with $N$ odd this cohomolgy group is trivial and we can safely ignore fractionalization.} The simplest example of a duality symmetry can be found in $\mathbb{Z}_2$ gauge theory, where the unique duality symmetry permutes $e$ and $m$. 

We will show that gauging a duality symmetry of a (2+1)D $\mathbb{Z}_N\times\mathbb{Z}_N$ gauge theory gives $\mathcal{Z}[\mathrm{TY}_{\mathbb{Z}_N\times\mathbb{Z}_N}^{\chi,\epsilon}]$. In this construction, the bicharacter $\chi$ is determined by the choice of anyon permutation and $\epsilon$ is determined by whether or not we stack a $(2+1)D$ $\mathbb{Z}_2$ SPT before gauging. Specifically, $H^3(\mathbb{Z}_2,U(1))=\mathbb{Z}_2$, so there is a nontrivial $\mathbb{Z}_2$ SPT in $(2+1)D$, in addition to the trivial theory \cite{chen2011,levin2012}. Because the symmetry fluxes of this SPT are semions, with topological spin $i$, stacking with this SPT changes the topological spin of the duality defects by $i$. We will show this explicitly in Appendix~\ref{sstringnet}.

Consider the 0-form duality symmetry with the permutation action
\begin{equation}\label{diagduality}
e_1\leftrightarrow m_1\qquad e_2\leftrightarrow m_2.
\end{equation}
This $\mathbb{Z}_{2}$ 0-form symmetry leads to twisted sector duality defects which are topological lines in spacetime (equivalently points in space) that can end the $\mathbb{Z}_{2}$ 0-form symmetry (see \eqref{defectbraid}). To characterize this particular duality symmetry in $\mathbb{Z}_N\times\mathbb{Z}_N$ gauge theory it is convenient to generate all the anyons in the theory with $\{m_1e_1^{-1},m_2e_2^{-1},m_1,m_2\}$ rather than the usual basis $\{m_1,e_1,m_2,e_2\}.$  
There are $N^2$ different flavors of duality defects, labeled by
\begin{equation}\label{sigmaflavors}
\sigma^{(p,q)}=(m_1^pm_2^q,\sigma).
\end{equation}
These flavors of $\sigma$ differ because they braid differently with $m_1e_1$ and $m_2e_2$. Specifically, $m_1e_1$ and $m_2e_2$ pass through the symmetry defect line unchanged, and can move around the duality defect $\sigma^{(p,q)}$ at the end of the defect line:
\begin{equation}\label{defectbraid}
\includegraphics{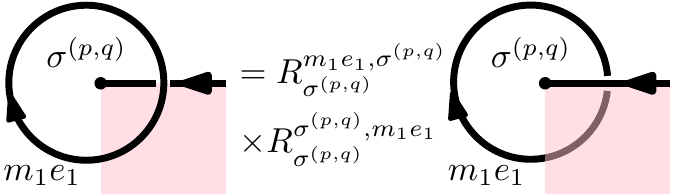},
\end{equation}
Here, the pink region indicates the permutation symmetry action on one side of the defect branch cut line. This permutation acts trivially on $m_1e_1$.
A full braid gives a phase $R^{m_1e_1,\sigma^{(p,q)}}_{\sigma^{(p,q)}}R^{\sigma^{(p,q)},m_1e_1}_{\sigma^{(p,q)}}$, which depends on $(p,q)$. Physically, these duality defects can be thought of as differing by a $m_1^pm_2^q$ anyon sitting next to a bare $\sigma$. 

The abelian anyons and the $N^2$ duality defects form a $\mathrm{Vec}_{\mathbb{Z}_N\times\mathbb{Z}_N}^{1}\times\mathrm{TY}_{\mathbb{Z}_N\times\mathbb{Z}_N}^{\chi,\epsilon}$ fusion category. Specifically, the anyons generated $\{m_1e_1^{-1},m_2e_2^{-1}\}$ together with $\sigma^{(0,0)}$ form a $\mathrm{TY}_{\mathbb{Z}_N\times\mathbb{Z}_N}^{\chi,\epsilon}$ fusion category and the anyons generated by $\{m_1,m_2\}$ form a $\mathrm{Vec}_{\mathbb{Z}_N\times\mathbb{Z}_N}^{1}$ fusion category. A similar analysis applies to any other duality symmetry, with any permutation action satisfying (\ref{permutetheta}).

\subsubsection{Gauging the duality symmetry}\label{sgaugeduality}

Now we gauge the duality symmetry to produce the center $\mathcal{Z}[\mathrm{TY}_{G}^{\chi,\epsilon}]$. Anyons in the gauged theory are labeled by a pair $([a],\pi_a)$ \cite{barkeshli2019}, where $a$ ranges over objects in the original category and twisted sectors. Here, $[a]$ is the orbit of the object $a$ under the duality symmetry, while $\pi_a$ is an irreducible representation of the stabilizer of $a$.  Since the symmetry group we are gauging is simply a $\mathbb{Z}_{2}$ the stabilizer is either $\mathbb{Z}_{2}$ or trivial depending on whether the anyon $a$ is duality invariant or not.\footnote{For gauging more general symmetries, we must consider projective representations of the stabilizer.  We do not encounter these here for the same reason that we do not find symmetry fractionalization.  See footnote \ref{foot1}.} Anyons in the gauged theory are grouped into orbits since after gauging those that differ by duality are physically equivalent.  Meanwhile the additional label of $\pi_{a}$ accounts for charges of the new $\mathbb{Z}_{2}$ duality gauge fields.

We will analyze the $N=2$ case, and then we will analyze the $N>2,N$ odd case. In $\mathbb{Z}_2\times\mathbb{Z}_2$ gauge theory enriched with diagonal $e-m$ duality symmetry given by (\ref{diagduality}), all the $\mathbb{Z}_2$ orbits are singletons except 
\begin{equation}\label{orbitz2}
[m_1],[m_2],[m_1m_2],[e_1m_2],[m_1e_1m_2],[m_1m_2e_2],
\end{equation}
where $[m_1]=\{m_1,e_1\}$, etc. The stabilizer subgroup of the singletons is $\mathbb{Z}_2$ and that of the other orbits (\ref{orbitz2}) is $\mathbb{Z}_1$. Every singleton can carry an irreducible representation of $\mathbb{Z}_2$, which we can label $r\in\mathbb{Z}_2=\{0,1\}$. Therefore, the gauged theory has anyons 
\begin{align}
\begin{split}\label{gaugedanyons}
&(1,r), (e_1m_1,r), (e_2m_2,r),(e_1m_1e_2m_2,r), \\
& [m_1],[m_2],[m_1m_2],[e_1m_2],[m_1e_1m_2],[m_1m_2e_2]\\
&(\sigma^{(0,0)},r), (\sigma^{(1,0)},r),(\sigma^{(0,1)},r), (\sigma^{(1,1)},r).
\end{split}
\end{align}

In summary, there are eight abelian anyons (first line), six non-abelian anyons from the orbits of size two (second line), and eight non-abelian anyons from the duality defects (third line), matching the anyon content of $\mathcal{Z}[\mathrm{TY}_{\mathbb{Z}_2\times\mathbb{Z}_2}^{\chi,\epsilon}]$ \cite{gelaki2009,evans2020}.

The topological spins of these anyons are given by \cite{barkeshli2019}
\begin{equation}\label{topspin}
\theta_{([a],\pi_a)}=\theta_{a_{\mathbf{g}}}\frac{\chi_{\pi_a}(\mathbf{g})}{\chi_{\pi_a}(1)},
\end{equation}
where $\chi_{\pi_a}(\mathbf{g})=\mathrm{Tr}[\pi_a(\mathbf{g})]$. Roughly speaking, \eqref{topspin} comes from the fact that the anyons in the gauged theory are dyons whose topological spin comes from both the original spin of $a_{\mathbf{g}}$ (first factor) along with the Aharonov Bohm phase from internal braiding of the anyon's flux and charge (second factor). The second factor is only relevant for the singletons under the symmetry. Note that the anyons of the underlying abelian topological order (i.e.\ the theory before gauging) are always in the untwisted sector ($\mathbf{g}$ trivial) so the second factor in \eqref{topspin} is equal to $1$ for the abelian anyons. Thus, the abelian anyons of the gauged theory come in pairs with identical spin. On the other hand, the duality defects are in the twisted sector ($\mathbf{g}$ non-trivial), so the resulting duality anyons come in pairs with spin differing by a minus sign. 

From (\ref{topspin}), we see that gauging the duality symmetry given in (\ref{diagduality}) gives a topological order with four abelian anyons with topological spin 1 and four abelian anyons with topological spin $-1$. The topological spins of the anyons coming from orbits of size two can be read off from (\ref{gaugedanyons}): there are four bosons and two fermions. In Appendix~\ref{sstringnet} (see Eqs.~\ref{selfstatabelian} and \ref{spinbgh}), we show that the abelian anyons and anyons with quantum dimension two can be labeled by $a_g^{(r)}$  and $b_{g,h}$ respectively, where $g,h\in G$ and $g\neq h$. The topological spins of these anyons are given by
\begin{equation}\label{topspinab}
\theta_{a_g^{(r)}}=\frac{1}{\chi(g,g)}\qquad\theta_{b_{g,h}}=\frac{1}{\chi(g,h)}.
\end{equation}

Therefore, the above analysis means that gauging this particular duality symmetry gives $\mathcal{Z}[\mathrm{TY}_{\mathbb{Z}_2\times\mathbb{Z}_2}^{\chi,\epsilon}]$ with the diagonal bicharacter \cite{thorngrenwang1}, given by (\ref{diagxyz}).

Finally, we must also obtain the topological spins of the duality anyons. To do so, we must either use the heptagon equations to obtain the topological spins of the duality defects, and then apply (\ref{topspin}), or directly solve the string-net equations. We show in Appendix~\ref{sstringnet}, using the latter approach, that topological spins of the duality anyons are $\left\{\pm\sqrt{i\epsilon},\pm\sqrt{\epsilon},\pm\sqrt{\epsilon},\pm\sqrt{-i\epsilon}\right\}$. 

For general odd $N$, $\mathbb{Z}_N\times\mathbb{Z}_N$ gauge theory enriched with the diagonal duality symmetry in \eqref{diagduality} has $N^4$ abelian anyons, together with $N^2$ flavors of $\sigma$, as written in Eq.~\eqref{sigmaflavors}. Gauging the duality symmetry gives anyons which are again labeled by $([a],\pi_a)$. Anyons generated by $m_1e_1$ and $m_2e_2$, along with the duality defects, form singletons under the duality symmetry.  
Since the singletons can carry an irreducible representation of $\mathbb{Z}_2$, we label them by $\pi_a\in\mathbb{Z}_2$. This gives in total:
\begin{itemize}
\item{$2N^2$ abelian anyons $a_g^{(r)}$},
\item{$\frac{N^2(N^2-1)}{2}$ dimension two anyons $b_{g,h}\sim b_{h,g}$ ($g\neq h$)},
\item{$2N^2$ dimension $N$ duality anyons $c_g^{(r)}$},
\end{itemize}
which matches the anyon content of $\mathcal{Z}[\mathrm{TY}_{\mathbb{Z}_N\times\mathbb{Z}_N}^{\chi,\epsilon}]$ \cite{gelaki2009}. The topological spins of $a_{g}^{(r)}$ and $b_{g,h}$ are again given by (\ref{topspinab}), and we refer to Appendix~\ref{sstringnet} for a computation of the topological spins of the duality anyons.

Another choice for the permutation of the duality symmetry in $\mathbb{Z}_N\times\mathbb{Z}_N$ gauge theory is
\begin{equation}\label{offdiagbicharacter}
e_1\leftrightarrow m_2\qquad e_2\leftrightarrow m_1.
\end{equation}

In the $N=2$ case, following the same calculations as above, we find that gauging this duality symmetry gives $\mathcal{Z}[\mathrm{TY}_{\mathbb{Z}_2\times\mathbb{Z}_2}^{\chi,\epsilon}]$ with eight bosonic $a_g^{(r)}$, three bosonic $b_{g,h}$, and three fermionic $b_{g,h}$.  
It follows from these topological spins that gauging the above duality symmetry gives $\mathcal{Z}[\mathrm{TY}_{\mathbb{Z}_2\times\mathbb{Z}_2}^{\chi,\epsilon}]$ with the off-diagonal bicharacter \cite{thorngrenwang1}, given in (\ref{offdiagxyz}). We will show in Appendix~\ref{sstringnet} that the topological spins of the duality anyons $c_{g}^{(r)}$ in this case are $\{\pm\sqrt{\epsilon},\pm\sqrt{\epsilon},\pm\sqrt{\epsilon},\pm i\sqrt{\epsilon}\}$.

We also summarize the number of bosonic duality anyons in $\mathcal{Z}[\mathrm{TY}_{\mathbb{Z}_N\times\mathbb{Z}_N}^{\chi,\epsilon}]$ in Table~\ref{table:table1}. These results are derived in Appendices~\ref{sstringnet} and~\ref{ssum}, and will be important for the following analysis of anomalies.

\begin{table}[tb]
\centering
\begin{tabular}{ |c|c|c| } 
\hline
 & Diagonal $\chi$ & Off-diagonal $\chi$\\
\hline
$N=2,\epsilon=+1$ & 2 & 3 \\
\hline
$N=2,\epsilon=-1$ & 0 & 1 \\ 
\hline
$N>2,N$ odd, $\epsilon=+1$ & $2N-1^*$ & $2N-1$ \\
\hline
$N>2,N$ odd, $\epsilon=-1$ & $0^*$ & $0$ \\ 
\hline
\end{tabular}
\caption{The number of bosonic duality anyons of $\mathcal{Z}[\mathrm{TY}_{\mathbb{Z}_N\times\mathbb{Z}_N}^{\chi,\epsilon}]$ for various choices of $N$ and $\epsilon$, for the diagonal and off-diagonal bicharacters. The asterisk in the diagonal column indicates that we only compute the number for $N$ such that $-1$ is a quadratic residue mod $N$.}
\label{table:table1}
\end{table}

\subsection{Analysis of anomalies}\label{sungaugeanomaly}
Now that we have the topological spins of the anyons in $\mathcal{Z}[\mathrm{TY}_{\mathbb{Z}_N\times\mathbb{Z}_N}^{\chi,\epsilon}]$, we can proceed to study when the fusion category is anomalous. We will first discuss a 0th-level obstruction to the fusion category being anomaly-free, detected by the absence of a duality-invariant $\tilde{\mathcal{L}}_m$ of the $G$ gauge theory. This obstruction already shows that $\mathrm{TY}_{\mathbb{Z}_N}^{\chi,\epsilon}$ is anomalous regardless of the choice of $\chi$ and $\epsilon$ \cite{thorngrenwang1}. This matches with the familiar fact that the self-dual clock model, which generalizes the Ising spin chain to $\mathbb{Z}_N$ symmetry, must be gapless. Furthermore, this 0th level obstruction shows that $\mathrm{TY}_{\mathbb{Z}_N\times\mathbb{Z}_N}^{\chi,\epsilon}$ is anomalous when $\chi$ is the diagonal bicharacter and $-1$ is not a quadratic residue mod $N$. 

We thus show that the existence of a duality invariant $\tilde{\mathcal{L}}_m$ is equivalent to the condition obtained in \cite{tambara2000} that for $|G|$ odd, the bicharacter must be \emph{hyperbolic} (see below \eqref{hyperbolic}) for $\mathrm{TY}_{G}^{\chi,\epsilon}$ to be anomaly-free. We then study the duality symmetries in (\ref{diagduality}) and (\ref{offdiagbicharacter}) for the particular $N$ for which they give hyperbolic bicharacters. We show that for $N>2,N$ odd (and giving a hyperbolic bicharacter), $\epsilon=-1$ gives another obstruction, related to the fact that in this case none of the duality anyons $c_g^{(r)}$ are bosonic.
\subsubsection{Duality-invariant $\tilde{\mathcal{L}}_m$}\label{sinvariantgroup}
In general, for the $\mathrm{TY}_{G}^{\chi,\epsilon}$ to be non-anomalous, there must first exist a duality-invariant $\tilde{\mathcal{L}}_m$ of the original $G$-gauge theory. Otherwise, there is no $(1+1)D$ gapped phase symmetric under the abelian $G$ symmetry that has the duality symmetry at all, regardless of whether or not the duality symmetry is spontaneously broken \cite{thorngrenwang1,choi2022,apte2022}. In this case, the fusion category is anomalous without even specifying the $F$ symbol. All Lagrangian algebras of $\mathcal{Z}[\mathrm{TY}_G^{\chi,\epsilon}]$ for $\mathrm{TY}_{G}^{\chi,\epsilon}$ with this 0th level obstruction describe theories that not only spontaneously break part of the fusion category symmetry (which is expected of anomalous fusion categories), but also describe theories that spontaneously break a subgroup of the invertible $G$ part of the fusion category. 

Note that duality symmetry in $(1+1)D$ is a gauging operation, so the absence of a duality-invariant $\tilde{L}_m$ means that there is no $G$-symmetric $(1+1)D$ theory that is invariant under gauging: gauging always maps one phase to a different phase. Therefore, the first step is to find for which $G$ there exists at least one duality-invariant $\tilde{\mathcal{L}}_m$ of the $G$ gauge theory. Thus, the results in this section can all be derived by explicit gauging using the actions of the $(1+1)D$ theories, as shown in \cite{thorngrenwang1,choi2022,apte2022}.

A duality-invariant $\tilde{\mathcal{L}}_m$ is a Lagrangian subgroup that satisfies two properties. First, it does not contain any pure charges of the $G$ gauge theory. Second, it maps back to itself under the anyon permutation of the duality symmetry. 

Let us use the absence of a duality-invariant $\tilde{\mathcal{L}}_m$ to show that $\mathrm{TY}_{\mathbb{Z}_N}^{\chi,\epsilon}$ is always anomalous. To see why there is no duality-invariant $\tilde{\mathcal{L}}_m$ in untwisted $\mathbb{Z}_N$ gauge theory, note that the only $\tilde{\mathcal{L}}_m$ is the one generated by $m$: $\tilde{\mathcal{L}}_m=\{\mathbf{1},m,\dots m^{N-1}\}$. This follows from the observation in Sec.~\ref{sabelian} that such a Lagrangian subgroup must contain one element from each equivalence class under modding out by $e$. In particular, it must contain an anyon of the form $me^p$, but this is not a boson unless $p=0$, so the Lagrangian subgroup must be generated by $m$. However, it is clear that this Lagrangian subgroup is not duality-invariant, because it maps to $\tilde{\mathcal{L}}_e$. Therefore, there are no duality-invariant $\tilde{\mathcal{L}}_m$ in $\mathbb{Z}_N$ gauge theory.\footnote{There does exist a duality-invariant Lagrangian subgroup when $N$ is a perfect square, but these always intersect nontrivially with $\tilde{\mathcal{L}}_e$. For example, for $N=4$, the duality-invariant Lagrangian subgroup is $\{1,e^2,m^2,e^2m^2\}$. In general, these Lagrangian subgroups describe $(1+1)D$ theories where the $\mathbb{Z}_N$ symmetry is spontaneously broken to its $\mathbb{Z}_{\sqrt{N}}$ subgroup.}

The analysis becomes more interesting for $G=\mathbb{Z}_N\times\mathbb{Z}_N$, where there does exist duality-invariant $\tilde{\mathcal{L}}_m$. For the diagonal duality symmetry (\ref{diagduality}), for $N=2$, the duality-invariant $\tilde{\mathcal{L}}_m$ is 
\begin{equation}\label{lminvariant}
\tilde{\mathcal{L}}_m=\{1,m_1e_2,m_2e_1,m_1e_2m_2e_1\}.
\end{equation}

This also serves as a duality invariant $\tilde{\mathcal{L}}_m$ for the off-diagonal duality symmetry (\ref{offdiagbicharacter}).

To determine for which $N$ there exists such a duality-invariant Lagrangian subgroup for $N>2$, we can generate the Lagrangian subgroup with 
\begin{equation}
\{m_1e_2^{p},m_2e_1^{-p}\}.
\end{equation}

As discussed in Sec.~\ref{sspt1d}, all magnetic Lagrangian subgroups of $\mathbb{Z}_N\times\mathbb{Z}_N$ gauge theory are of this form. For the diagonal duality, we need this Lagrangian subgroup to be invariant under $m_1\leftrightarrow e_1$ and $m_2\leftrightarrow e_2$. This means that $m_2e_1^{-p}$ must generate $e_1m_2^{p}$ and $m_1e_2^{p}$ must generate $e_2m_1^{-p}$, so we have
\begin{align}
\begin{split}
&\left(m_2e_1^{-p}\right)^p=m_2^pe_1^{-p^2}=m_2^pe_1\\
&\left(m_1e_2^p\right)^{-p}=m_1^{-p}e_2^{-p^2}=e_2m_1^{-p},
\end{split}
\end{align}
which means that
\begin{equation}\label{p2cond}
p^2=-1\qquad\text{mod }N.
\end{equation}
The existence of such a $p$ is a constraint on $N$, which implies that all prime factors of $N$ are 1 mod 4, with the exception of at most a single factor of 2.  Because there only exists a duality-invariant Lagrangian subgroup for $N$ where there exists $p$ satisfying (\ref{p2cond}), only for these $N$ can $\mathrm{TY}_{\mathbb{Z}_N\times\mathbb{Z}_N}^{\chi,\epsilon}$ be anomaly-free \cite{choi2022,apte2022}. For example, for $N=2$, $p=1$ satisfies this condition, and for $N=5$, $p=2$ and $p=3$ satisfy this condition. For any $N$, if there exists a $p$ that satisfies (\ref{p2cond}), $N-p$ also satisfies (\ref{p2cond}). 

For the off-diagonal duality, we only require that $m_1e_2^p$ generates $e_2m_1^p$ and $m_2e_1^{-p}$ generates $e_1m_2^{-p}$. This means that
\begin{align}
\begin{split}
&\left(m_1e_2^p\right)^p=m_1^pe_2^{p^2}=m_1^pe_1\\
&\left(m_2e_1^{-p}\right)^{-p}=m_2^{-p}e_2^{p^2}=e_1m_2^{-p},
\end{split}
\end{align}
which is satisfied when 
\begin{equation}\label{p21N}
p^2=1\qquad\text{mod }N.
\end{equation}

For any $N$, $p=1$ is always a solution to (\ref{p21N}), so $\mathrm{TY}_{\mathbb{Z}_N\times\mathbb{Z}_N}^{\chi,\epsilon}$ can be anomaly-free for any $N$.

More generally, it was shown in Ref.~\onlinecite{tambara2000} that $\mathrm{TY}_{G}^{\chi,\epsilon}$ is anomaly-free when $|G|$ is odd if and only if $\epsilon=+1$ and the bicharacter $\chi$ is \emph{hyperbolic}. This means that $G$ is of the form 
\begin{equation}\label{hyperbolic1}
G=G_1\times G_2,
\end{equation}
where $G_1\cong G_2$ and
\begin{equation}\label{hyperbolic}
\chi(g_1,g_1')=\chi(g_2,g_2')=1,
\end{equation}
for all $g_1,g_1'\in G_1$ and $g_2,g_2'\in G_2$.

Note that for $G=\mathbb{Z}_N\times\mathbb{Z}_N$, the diagonal bicharacter when $-1$ is a quadratic residue mod $N$ and the off-diagonal bicharacter for any odd $N$ both satisfy this condition. The off-diagonal bicharacter clearly satisfies the hyperbolic condition because $\chi((1,0),(1,0))=\chi((0,1),(0,1))=1$. The diagonal bicharacter where $-1$ is a quadratic residue mod $N$ also satisfies the hyperbolic condition because we can choose to generate $\mathbb{Z}_N\times\mathbb{Z}_N$ with $(1,p)$ and $(1,-p)$. Then we find that 
\begin{align}
\begin{split}
    \chi((1,p),(1,p))&=\chi((1,-p),(1,-p))\\
    &=e^{\frac{2\pi i }{N}}e^{\frac{2\pi ip^2}{N}},
\end{split}
\end{align}
which is equal to $1$ if and only if $p^2=-1$ mod $N$. Note that $(1,p)$ and $(1,-p)$ generate $\mathbb{Z}_N\times\mathbb{Z}_{N}$ in this case because $p$ is coprime to $N$ and $N$ is odd.

We show in Appendix~\ref{shyperbolic} that the condition that there exists a duality-invariant $\tilde{\mathcal{L}}_m$ of the $G$ gauge theory is actually equivalent to the condition that the bicharacter is hyperbolic. The intuition behind this result is the following: the bicharacter gives the topological spins of the anyons in $\mathcal{Z}[\mathrm{TY}_G^{\chi,\epsilon}]$ that come from anyons of the $G$ gauge theory (singletons and orbits of size two under the duality) according to (\ref{topspinab}). Therefore, the properties of the bicharacter are specified entirely by the $G$ gauge theory and the choice of permutation, like the existence of a duality invariant $\tilde{\mathcal{L}}_m$.
\subsubsection{Obstructions from counting bosons}\label{sungaugeanomaly}
We now specifically consider $G=\mathbb{Z}_N\times\mathbb{Z}_N$ with $N=2$ or $N>2$ and odd, for which there exists at least one duality-invariant $\tilde{\mathcal{L}}_m$. In particular, this means that for the diagonal bicharacter, we only consider $N=2$ and $N$ odd for which $-1$ is a quadratic residue mod $N$. For these $G$, any anomaly of the TY fusion category comes from its $F$ symbol. We will use the results in Table~\ref{table:table1} to show how TY fusion categories with certain $N,\chi,$ and $\epsilon$ are anomalous. We recover the fact that for $N=2$, the diagonal bicharacter with $\epsilon=-1$ is anomalous, and for $N>2,N$ odd, both the diagonal and the off-diagonal bicharacters with $\epsilon=-1$ are anomalous \cite{tambara2000}.

A simple calculation using the anyon data listed at the end of Sec.~\ref{sgaugeduality} shows that the total quantum dimension of $\mathcal{Z}[\mathrm{TY}_{\mathbb{Z}_N\times\mathbb{Z}_N}^{\chi,\epsilon}]$ is 
\begin{equation}
\mathcal{D}=2N^2.
\end{equation}

According to (\ref{dim}), this is the quantum dimension of any Lagrangian algebra $\mathcal{L}$ of $\mathcal{Z}[\mathrm{TY}_{\mathbb{Z}_N\times\mathbb{Z}_N}^{\chi,\epsilon}]$. At the very least, we need a sufficient number of bosons $a$ (other than the ones in $\mathcal{L}_e$) to form
\begin{equation}
\mathcal{L}_m=\oplus_an_aa,
\end{equation}
where $\sum_an_ad_a=2N^2$. Of course, if we have any bosons at all in $\mathcal{Z}[\mathrm{TY}_{\mathbb{Z}_N\times\mathbb{Z}_N}^{\chi,\epsilon}]$, we can obtain $\sum_an_ad_a=2N^2$ by choosing sufficiently large $\{n_a\}$. However, there are constraints on $\{n_a\}$ given by the fact that $n_{\mathbf{1}}=1$ and the inequality in (\ref{ineq}). We will use these constraints to demonstrate that certain TY fusion categories are anomalous because their corresponding Drinfeld center does not have enough bosons. 

First, note that for the abelian anyons, $n_a\in\{0,1\}$ due to (\ref{ineq}). Specifically, if the abelian anyon $a$ is in $\mathcal{L}$, then $\bar{a}$ is also in $\mathcal{L}$. However from (\ref{ineq}) and $n_{\mathbf{1}}=1$, we have
\begin{equation}
n_an_{\bar{a}}\leq 1\to n_a=n_{\bar{a}}\in\{0,1\}
\end{equation}

To further analyze possible obstructions, let us first focus on $N=2$, with the diagonal bicharacter given in (\ref{diagxyz}). We show in Appendix~\ref{sstringnet} that for $\epsilon=+1$, there are the following bosons:
\begin{itemize}
\item{$a_{(0,0)}^{(0)},a_{(0,0)}^{(1)},a_{(1,1)}^{(0)},a_{(1,1)}^{(1)}$}
\item{$b_{(0,0),(1,0)},b_{(0,0),(0,1)},b_{(0,0),(1,1)},b_{(1,0),(0,1)}$}
\item{$c_{(1,0)}^{(0)},c_{(0,1)}^{(0)}$}
\end{itemize}

On the other hand for $\epsilon=-1$, there are no bosonic duality anyons according to Table~\ref{table:table1}.

Notice that five of these anyons already belong in $\mathcal{L}_e$, which is given by\footnote{We can obtain $\mathcal{L}_e$ from the string-net construction, as shown in Appendix~\ref{sstringnet}. Alternatively, we can also note that $\mathcal{L}_e$ comes from $\tilde{\mathcal{L}}_e$ of $\mathbb{Z}_2\times\mathbb{Z}_2$ gauge theory: $\tilde{\mathcal{L}}_e=\{\mathbf{1},e_1,e_2,e_1e_2\}$. $e_1,e_2$, and $e_1e_2$ become dimension two anyons in $\mathcal{Z}[\mathrm{TY}_{\mathbb{Z}_2\times\mathbb{Z}_2}^{\chi,\epsilon}]$ because they form orbits of size two under the duality symmetry. $\mathbf{1}$ becomes $a_{(0,0)}^{(0)}$ and $a_{(0,0)}^{(1)}$ in $\mathcal{Z}[\mathrm{TY}_{\mathbb{Z}_2\times\mathbb{Z}_2}^{\chi,\epsilon}]$.} 
\begin{align}
\begin{split}
\mathcal{L}_e&=a_{(0,0)}^{(0)}+a_{(0,0)}^{(1)}+b_{(0,0),(0,1)}\\
&+b_{(0,0),(1,0)}+b_{(0,0),(1,1)}.
\end{split}
\end{align}

Now we use the fusion rule
\begin{align}
\begin{split}
b_{(1,0),(0,1)}\times &b_{(1,0),(0,1)}\\
&=a_{(0,0)}^{(0)}+a_{(0,0)}^{(1)}+a_{(1,1)}^{(0)}+a_{(1,1)}^{(1)},
\end{split}
\end{align}
which can be derived from the duality-enriched gauge theory \cite{barkeshli2019}. Combining this with (\ref{ineq}) and the fact that $n_{a_{(0,0)}^{(1)}}=0$ in $\mathcal{L}_m$ (because it is in $\mathcal{L}_e$), we find that
\begin{equation}
n_{b_{(1,0),(0,1)}}^2\leq 3
\end{equation}
in $\mathcal{L}_m$, so $n_{b_{(1,0),(0,1)}}\in\{0,1\}$. This means that we need at least one duality anyon to be bosonic, because using the anyons described thus far we can only construct a Lagrangian algebra with dimension $3+2*1=5$, which is less than $2N^2=8$. However, none of the duality anyons are bosonic if $\epsilon=-1$, so $\mathrm{TY}_{\mathbb{Z}_2\times\mathbb{Z}_2}^{\chi,\epsilon}$ with diagonal $\chi$ and $\epsilon=-1$ must be anomalous. On the other hand, for $\epsilon=+1$, there is no obstruction to there being a $\mathcal{L}_m$ at the level of number of bosons, because there are two bosonic duality anyons. Indeed, this fusion category is not anomalous \cite{tambara2000,thorngrenwang1}. Performing a similar analysis with the off-diagonal bicharacter, we find that there is no obstruction (at the level of number of bosons) to there being a $\mathcal{L}_m$ for both $\epsilon=+1$ and $\epsilon=-1$, matching the result of \cite{tambara2000,thorngrenwang1}.

It follows from arguments similar to the $N=2$ cause that for $N>2$, we need $\mathcal{Z}[\mathrm{TY}_{\mathbb{Z}_N\times\mathbb{Z}_N}^{\chi,\epsilon}]$ to have at least one bosonic duality anyon to obtain a $\mathcal{L}_m$. For the diagonal bicharacter and $N$ for which $-1$ is a quadratic residue mod $N$, the results stated in Table~\ref{table:table1} and proven in Appendices~\ref{sstringnet} and~\ref{ssum} show that there are $2N-1$ bosonic duality anyons when $\epsilon=+1$ and no bosonic duality anyons when $\epsilon=-1$. Therefore, there is no obstruction to there being a $\mathcal{L}_m$ when $\epsilon=+1$, but the case with $\epsilon=-1$ is always anomalous. There is a similar story for the off-diagonal bicharacter: $\mathrm{TY}_{\mathbb{Z}_N\times\mathbb{Z}_N}^{\chi,\epsilon}$ with $\epsilon=+1$ can be anomaly-free (for any odd $N$). However, all such fusion categories with $\epsilon=-1$ are anomalous because there are no bosonic duality anyons in $\mathcal{Z}[\mathrm{TY}_{\mathbb{Z}_N\times\mathbb{Z}_N}^{\chi,\epsilon}]$.
\section{Discussion}\label{sdiscussion}
In this work, we presented a general approach for understanding anomalies of fusion category symmetries, and applied it to study anomalies of TY fusion categories. This yields highly computable obstructions to the fusion category being anomaly-free. We highlight here some future directions.

First, it would be instructive to understand better the mapping between module categories and Lagrangian algebras of the corresponding Drinfeld center, especially from a physical perspective. This would also lead to a direct mapping between fiber functors and magnetic Lagrangian algebras. A starting point for this problem is the case when the fusion category is $\mathrm{Rep}_G$. Here, both module categories over $\mathrm{Rep}_G$ and Lagrangian algebras of $\mathcal{Z}[\mathrm{Rep}_G]$ correspond to gapped $(1+1)D$ theories with the $G$ symmetry, which are labeled by subgroups $K$ of $G$ and cocycles in $H^2(K,U(1))$ \cite{beigi2011}. 

The mapping envisioned above would also be helpful for constructing microscopic $(1+1)D$ lattice models corresponding to gapped phases labeled by Lagrangian algebras. In particular, using \cite{kitaev2012}, this would give lattice models for SPTs of fusion category symmetries when given an $\mathcal{L}_m$. This idea is also closely related to the problems of (1) constructing microscopic models for string-net models with boundaries, given a Lagrangian algebra, and (2) condensing a Lagrangian algebra microscopically, by adding terms to the lattice Hamiltonian. 

More generally, it would interesting to further explore the phase diagrams of $(1+1)D$ systems with fusion category symmetry, and transitions between different phases with fusion category symmetry. For example, for $\mathcal{Z}[\mathrm{TY}_{\mathbb{Z}_2\times\mathbb{Z}_2}^{\chi,\epsilon}]$ with off-diagonal bicharacter and $\epsilon=+1$, there are actually three distinct SPTs of the fusion category symmetry, but it is unclear how to drive a system into each of these gapped phases at either the field theory level or the lattice level.\footnote{For this particular fusion category, it is perhaps useful to notice that $\mathcal{Z}[\mathrm{TY}_{\mathbb{Z}_2\times\mathbb{Z}_2}^{\chi,\epsilon}]$ is dual to $\mathcal{Z}[\mathrm{Rep}_{D_4}]$ where the dihedral group $D_4$ is the symmetry group of a square.} 

It would also be interesting to apply our method to other fusion categories. As discussed in \eqref{qdanom}, one requirement on fusion categories that come from the fiber functor description of anomalies is that a fusion category can only be anomaly-free if the quantum dimensions of all its objects are integer \cite{chang2019}. It would be instructive to rederive this result using magnetic Lagrangian algebras, and study other fusion categories containing only objects with integer quantum dimension.  Relatedly, our analysis uses only coarse obstuctions based on the number of bosons in the corresponding Drinfeld center, and a natural direction for future work is to study other kinds of obstructions to the existence of a $\mathcal{L}_m$. 

Another natural direction for future work is to explore anomalies of non-invertible symmetries in higher dimensions \cite{choi2022,apte2022}, perhaps extending recent results on gapped boundaries of $(3+1)D$ topological orders \cite{zhao2022,ji2022,luo2022}.

Finally, in the group-like case, we not only know when the symmetry is anomalous but we also understand a classification of different kinds of anomalies. For example, for finite abelian groups, the types of anomalies $\omega\in H^3(G,U(1))$ are differentiated into type-I, type-II, and type-III, which are associated with three different kinds of braiding processes in $\mathcal{Z}[\mathrm{Vec}_G^{\omega}]$\cite{wang2015}. It would be interesting to further develop this kind of classification for anomalies of non-abelian groups and general fusion category symmetries. These different types are closely related to the linking invariants described in \cite{kaidi2023}.

\let\oldaddcontentsline\addcontentsline
\renewcommand{\addcontentsline}[3]{}
\section*{Acknowledgements}
\let\addcontentsline\oldaddcontentsline

We thank Michael Levin and Nat Tantivasadakarn for many helpful conversations, Michael Levin for comments on the manuscript, and Arkya Chatterjee, Meng Cheng, Kantaro Ohmori, Sal Pace, and Ryan Thorngren for related discussions. C.Z. is supported by the University of Chicago Bloomenthal Fellowship and the National Science Foundation Graduate Research Fellowship under Grant No. 1746045.  C.C. is supported by the US Department of Energy DE-SC0021432 and the Simons Collaboration on Global Categorical Symmetries. This research was supported in part by Perimeter Institute for Theoretical Physics. Research at Perimeter Institute is supported by the Government of Canada through the Department of Innovation, Science, and Economic Development and by the Province of Ontario through the Ministry of Research and Innovation.
\appendix
\section{$\mathcal{Z}[\mathrm{TY}_{\mathbb{Z}_N\times\mathbb{Z}_N}^{\chi,\epsilon}]$ from the string-net construction}\label{sstringnet}
The string-net construction is one way to obtain information about $\mathcal{Z}[\mathcal{A}]$ for any input fusion category $\mathcal{A}$\cite{levin2005,lin2021}. We will use this method to obtain some of the information relevant for identifying the Lagrangian algebras of $\mathcal{Z}[\mathrm{TY}_{\mathbb{Z}_N\times\mathbb{Z}_N}^{\chi,\epsilon}]$, including the topological spins of the anyons. 

The input fusion category gives a list of string types, which we label by English letters $\{a,b,c\dots\}$ to match with the notation of Ref.~\cite{lin2021} (note that this differs from the main text, where we use English letters to refer to anyons). For example, $\mathrm{TY}_{G}^{\epsilon,\chi}$ has $|G|+1$ string types corresponding to the $|G|$ group elements and the duality object. Each anyon $\alpha$ of $\mathcal{Z}[\mathrm{TY}_{G}^{\epsilon,\chi}]$ is characterized by a set of data $\left(\Omega_{\alpha},\bar{\Omega}_{\alpha},n_{\alpha}\right)$ where $n_{\alpha}=\left(n_{\alpha,g_1},\cdots,n_{\alpha,g_{|G|}},n_{\alpha,\sigma}\right)$ is a vector with $|G|+1$ elements, describing which strings $\alpha$ is built out of. The main equations for the string-net models are\cite{lin2021}:
\begin{align}
\begin{split}\label{maineq}
\sum_{a'}\Omega_{\alpha}^{a,rsa'}\left(F_{c'a'c}^{rab}\right)^*F_{c'a'b'}^{asb}&=\sum_t\Omega_{\alpha}^{c,rtc'}\bar{\Omega}_{\alpha}^{b,tsb'}F_{c'cb'}^{abt}\\
\bar{\Omega}_{\alpha}^{a,rsa'}&=\left(\Omega_{\alpha}^{a,sra'}\right)^*\\
\sum_s\bar{\Omega}_{\alpha}^{a,rsa'}\Omega_{\alpha}^{a,sta'}&=\delta_{rt},
\end{split}
\end{align}
where the $F$ symbol is given by (\ref{fsymbol}) and $\Omega$ describes the braiding of an anyon type $\alpha$ with the string type $a$\footnote{Here we use the same convention as Ref.~\onlinecite{lin2021}, where the string types are always oriented in the upward direction.}:
\begin{equation}
\includegraphics{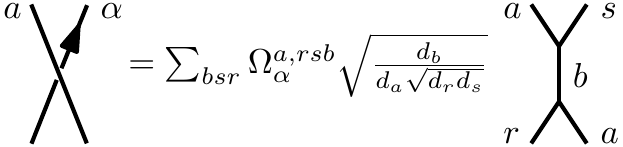}.
\end{equation}

In this work we will assume that $n_{\alpha,r},n_{\alpha,s}\in[0,1]$, which is sufficient for the string-net models we will consider. When $n_{\alpha,r}>1$ or $n_{\alpha,2}>1$ there are additional indices on $\Omega$\cite{lin2021}. 

The complete set of anyons describing $\mathcal{Z}[\mathcal{A}]$ correspond to the independent solutions to (\ref{maineq}) for $\left(\Omega_{\alpha},\bar{\Omega}_{\alpha},n_{\alpha}\right)$. For the particular case where $\mathcal{A}=\mathrm{TY}_{G}^{\chi,\epsilon}$, we will show that $\mathcal{Z}\left[\mathrm{TY}_{G}^{\chi,\epsilon}\right]$ has $\frac{7}{2}|G|+\frac{1}{2}|G|^2$ anyons, matching Ref.~\onlinecite{gelaki2009} and our derivation in Sec.~\ref{sgaugeduality}. These consist of:
\begin{itemize}
\item{$2|G|$ abelian anyons $a_g^{(0)}$ and $a_g^{(1)}$, where $g\in G$, built out of a single string type $g\in G$.}
\item{$\frac{|G|\left(|G|-1\right)}{2}$ non-abelian anyons $b_{g,h}$, with quantum dimension 2, built out of two string types $g,h\in G$, with $g\neq h$.}
\item{$2|G|$ non-abelian anyons $c_g^{(0)}$ and $c_g^{(1)}$, where $g\in G$, with quantum dimension $\sqrt{|G|}$, built out of a single string type $\sigma$.}
\end{itemize}

Note that in the string-net construction, the anyons in $\mathcal{L}_e$ are easy to specify: they are the anyons that contain the vacuum string type, $1$. For $\mathrm{TY}_{G}^{\chi,\epsilon}$, there are two abelian anyons and $|G|-1$ dimension two anyons that contain the vacuum string. 

To study the other Lagrangian algebras of $\mathcal{Z}[\mathrm{TY}_{\mathbb{Z}_N\times\mathbb{Z}_N}^{\chi,\epsilon}]$, we will need the self statistics and mutual statistics of the anyons. The self statistics of an anyon $\alpha$ is given by
\begin{equation}\label{selfstatformula}
e^{i\theta_{\alpha}}=\frac{\sum_s\mathrm{Tr}\left(\Omega_{\alpha}^{\bar{s},ss1}\right)d_s}{\sum_sn_{\alpha,s}d_s},
\end{equation}

We will also make use of the following formula for the $S$ matrix elements:
\begin{equation}\label{smatrix}
S_{\alpha\beta}=\frac{1}{D}\sum_{stb}\mathrm{Tr}\left(\bar{\Omega}_{\alpha}^{t,ssb}\right)\mathrm{Tr}\left(\bar{\Omega}_{\beta}^{s,ttb}\right)d_b.
\end{equation}

In particular, we will use (\ref{smatrix}) to propose $\mathcal{L}_m$ for $\mathrm{TY}_{\mathbb{Z}_2\times\mathbb{Z}_2}^{\chi,\epsilon}$ when the fusion category is not anomalous.
\subsection{Anyon self statistics}\label{sself}

We begin with computing the self statistics of the abelian anyons $a_g^{(0)}$ and $a_g^{(1)}$. Plugging in $a=\bar{g},r=s=t=g$, and $b=\sigma$, it follows from the fusion rules that $a'=1$ and $c'=c=b'=\sigma$. Then (\ref{maineq}) says that
\begin{align}
\begin{split}\label{abelian1}
\Omega^{\bar{g},gg1}_{a_g^{(r)}}\left(F^{g\bar{g}\sigma}_{\sigma 1\sigma}\right)^*F^{\bar{g}g\sigma}_{\sigma 1\sigma}&=\Omega^{\sigma,gg\sigma}_{a_g^{(r)}}\bar{\Omega}^{\sigma,gg\sigma}_{a_g^{(r)}}F^{\bar{g}\sigma g}_{\sigma\sigma\sigma}\\
\to\Omega^{\bar{g},gg1}_{a_g^{(r)}}&=F^{\bar{g}\sigma g}_{\sigma\sigma\sigma}=\chi(\bar{g},g).
\end{split}
\end{align}

From (\ref{selfstatformula}), we have 
\begin{equation}\label{selfstatabelian}
e^{i\theta_{a_g^{(0)}}}=e^{i\theta_{a_g^{(1)}}}=\chi(\bar{g},g)=\frac{1}{\chi(g,g)}.
\end{equation}

$a_g^{(0)}$ and $a_g^{(1)}$ differ because $\Omega^{\sigma,gg\sigma}_{a_g^{(0)}}=-\Omega^{\sigma,gg\sigma}_{a_g^{(1)}}$. To see this, first note that switching $\bar{g}$ with $h$ in (\ref{abelian1}) gives 
\begin{equation}\label{chihg}
\Omega^{h,gg(hg)}_{a_g^{(r)}}=\chi(h,g).
\end{equation}

Then plugging in $a=a'=b=\sigma$ and $r=s=t=g$ into (\ref{maineq}) gives 
\begin{equation}
\Omega^{\sigma,gg\sigma}_{a_g^{(r)}}\left(F^{g\sigma\sigma}_{(cg)\sigma c}\right)^*F^{\sigma g\sigma}_{(cg)\sigma\sigma}=\Omega^{c,gg (cg)}_{a_g^{(r)}}\bar{\Omega}^{\sigma,gg\sigma}_{a_g^{(r)}}F^{\sigma\sigma g}_{(cg)c\sigma}.
\end{equation}

Simplifying using (\ref{chihg}), we obtain
\begin{align}
\begin{split}
\left(\Omega^{\sigma,gg\sigma}_{a_g^{(r)}}\right)^2=\chi(c,g)\chi(g,gc)^{-1}&=\frac{1}{\chi(g,g)}\\
\to \Omega^{\sigma,gg\sigma}_{a_g^{(r)}}&=\pm\frac{1}{\sqrt{\chi(g,g)}},
\end{split}
\end{align}
so we choose 
\begin{equation}\label{sigmaggsigma}
\Omega^{\sigma,gg\sigma}_{a_g^{(0)}}=\frac{1}{\sqrt{\chi(g,g)}}\qquad \Omega^{\sigma,gg\sigma}_{a_g^{(1)}}=-\frac{1}{\sqrt{\chi(g,g)}}.
\end{equation}
Note that (\ref{sigmaggsigma}) will be important for computing the braiding statistics between $a_g^{(0)},a_g^{(1)}$ and the duality anyons. We will use these braiding statistics to hypothesize Lagrangian algebras.

We now compute the self statistics of $b_{g,h}$. Plugging in $a=\bar{g},r=s=g,$ and $b=\sigma$ gives
\begin{equation}
\Omega^{\bar{g},gg1}_{b_{g,h}}\left(F^{g\bar{g}\sigma}_{\sigma 1\sigma}\right)^*F^{\bar{g}g\sigma}_{\sigma 1\sigma}=\sum_t\Omega^{\sigma,gt\sigma}_{b_{g,h}}\bar{\Omega}^{\sigma,tg\sigma}_{b_{g,h}}F^{\bar{g}\sigma t}_{\sigma\sigma\sigma},
\end{equation}
where the right hand side runs over $t=g,h$. From (\ref{maineq}), we obtain $\Omega^{\sigma,gg\sigma}_{b_{g,h}}=0$ and $\Omega^{\sigma,gh\sigma}_{b_{g,h}}\bar{\Omega}^{\sigma,hg\sigma}_{b_{g,h}}=1$, so
\begin{equation}
\Omega^{\bar{g},gg1}_{b_{g,h}}=F^{\bar{g}\sigma h}_{\sigma\sigma\sigma}=\chi(\bar{g},h).
\end{equation}
Then from (\ref{selfstatformula}) we get
\begin{align}
\begin{split}\label{spinbgh}
e^{\theta_{b_{g,h}}}&=\frac{1}{2}\left(\Omega^{\bar{g},gg1}_{b_{g,h}}+\Omega^{\bar{h},hh1}_{b_{g,h}}\right)\\
&=\frac{1}{2}\left(\chi(\bar{g},h)+\chi(\bar{h},g)\right)\\
&=\frac{1}{\chi(g,h)}.
\end{split}
\end{align}
Since these anyons are built from two string types $g$ and $h$ with $g\neq h$, there are $\begin{pmatrix} |G|\\ 2\end{pmatrix}=\frac{|G|(|G|-1)}{2}$ anyons of this kind.

Now we move onto the non-abelian anyons $c_g^{(0)}$ and $c_g^{(1)}$ with quantum dimension $\sqrt{|G|}$. Note that here, the subscript does not refer to the string type that it is built out of. Rather, we will define $g$ to be the following. Consider the generators $p_i$ of $G$, of order $|p_i|$, so that $\prod_i|p_i|=|G|$. We will show that
\begin{equation}\label{omegapi}
\left(\Omega^{p_i,\sigma\sigma\sigma}_{c_g^{(r)}}\right)^{|p_i|}=\frac{1}{\chi(p_i,p_i)^{\frac{|p_i|(|p_i|-1)}{2}}}.
\end{equation}
so that $\Omega^{p_i,\sigma\sigma\sigma}_{\alpha}$ can take $|p_i|$ different values that differ by a phase $e^{\frac{2\pi i}{|p_i|}}$. Therefore, there are $|G|$ different $\{\Omega^{p_i,\sigma\sigma\sigma}\}$, labeled by $g\in G$. Furthermore, for each $\{\Omega^{p_i,\sigma\sigma\sigma}\}$, we have two values of $\Omega^{\sigma,\sigma\sigma 0}_{\alpha}$ that differ by a sign, giving a total of $2|G|$ anyons of this kind.

To obtain the self statistics of these anyons, we will use the following two results:
\begin{equation}\label{result1}
\Omega^{g,\sigma\sigma\sigma}_{c_g^{(r)}}=\chi(g,gh)\Omega^{\sigma,\sigma\sigma(gh)}_{c_g^{(r)}}\bar{\Omega}^{\sigma,\sigma\sigma h}_{c_g^{(r)}},
\end{equation}
\begin{equation}\label{result2}
\Omega^{g,\sigma\sigma\sigma}_{c_g^{(r)}}\Omega^{h,\sigma\sigma\sigma}_{c_g^{(r)}}=\frac{1}{\chi(g,h)}\Omega^{(gh),\sigma\sigma\sigma}_{c_g^{(r)}}.
\end{equation}

In particular, plugging in $h=\bar{g}$ into (\ref{result1}) gives
\begin{equation}\label{gbarg}
\Omega^{g,\sigma\sigma\sigma}_{c_g^{(r)}}=\Omega^{\sigma,\sigma\sigma 1}_{c_g^{(r)}}\bar{\Omega}^{\sigma,\sigma\sigma\bar{g}}_{c_g^{(r)}}.
\end{equation}

For $g=1$, we have
\begin{equation}\label{omega0}
\Omega^{1,\sigma\sigma\sigma}_{c_g^{(r)}}=1.
\end{equation}

We derive (\ref{result1}) by plugging in $a=g,b'=h,$ and $r=s=b=t=\sigma$ into (\ref{maineq}) and we derive (\ref{result2}) by plugging in $r=s=t=\sigma$ and $a=g,b=h$ into (\ref{maineq}). Choosing $g=p_i$ and iterating (\ref{result2}) $|p_i|-1$ times and then using (\ref{omega0}), we obtain (\ref{omegapi}). We can get $\Omega^{g,\sigma\sigma\sigma}_{c_g^{(r)}}$ for all $g$ that are not generators of $G$ by using (\ref{result2}). This gives $\{\Omega^{g,\sigma\sigma\sigma}_{c_g^{(r)}}\}$.

Now we use $a=r=s=t=b=\sigma$ and $a'=g,c=b'=1$ in (\ref{maineq}) to get
\begin{equation}
\sum_{g}\Omega^{\sigma,\sigma\sigma g}_{c_g^{(r)}}\left(F^{\sigma\sigma\sigma}_{\sigma g1}\right)^*F^{\sigma\sigma\sigma}_{\sigma g1}=\Omega^{1,\sigma\sigma\sigma}_{c_g^{(r)}}\bar{\Omega}^{\sigma,\sigma\sigma 1}_{c_g^{(r)}}F^{\sigma\sigma\sigma}_{\sigma 11}.
\end{equation}

Using (\ref{omega0}) and (\ref{fsymbol}), we obtain
\begin{equation}
\frac{\epsilon^2}{|G|}\sum_g\Omega^{\sigma,\sigma\sigma g}_{c_g^{(r)}}=\bar{\Omega}^{\sigma,\sigma\sigma 1}_{c_g^{(r)}}\frac{\epsilon}{\sqrt{|G|}}.
\end{equation}
Further simplifying, and using (\ref{gbarg}), we obtain
\begin{align}
\begin{split}
\left(\bar{\Omega}^{\sigma,\sigma\sigma 1}_{c_g^{(r)}}\right)^2&=\frac{\epsilon}{\sqrt{|G|}}\sum_g\left(\Omega^{\bar{g},\sigma\sigma\sigma}_{c_g^{(r)}}\right)^*\\
&=\frac{\epsilon}{\sqrt{|G|}}\sum_g\left(\Omega^{g,\sigma\sigma\sigma}_{c_g^{(r)}}\right)^*.
\end{split}
\end{align}
It follows from (\ref{selfstatformula}) that
\begin{align}
\begin{split}\label{nonabelianstat}
e^{i\theta_{c_g^{(0)}}}&=\sqrt{\frac{\epsilon}{\sqrt{|G|}}\sum\Omega^{g,\sigma\sigma\sigma}_{\alpha}}\\
 e^{i\theta_{c_g^{(1)}}}&=-\sqrt{\frac{\epsilon}{\sqrt{|G|}}\sum\Omega^{g,\sigma\sigma\sigma}_{\alpha}},
\end{split}
\end{align}
where $\{\Omega^{g,\sigma\sigma\sigma}_{\alpha}\}$ are given by (\ref{omegapi}) and (\ref{result2}).

These computations match with Sec. 4C of \cite{gelaki2009}.

\subsection{Example: $G=\mathbb{Z}_2\times\mathbb{Z}_2$}\label{sexample}
We will now consider conditions for the existence of a $\mathcal{L}_m$ for a TY fusion category with $G=\mathbb{Z}_2\times\mathbb{Z}_2$. For this group, there are four different TY fusion categories, given by two possible bicharacters and two values of $\epsilon$. First, let's consider the diagonal bicharacter given in (\ref{diagxyz}). $\mathcal{L}_e$ is given by all the anyons that contain the trivial string:
\begin{align}
\begin{split}
\mathcal{L}_e&=a_{(0,0)}^{(0)}+a_{(0,0)}^{(1)}+b_{(0,0),(1,0)}\\
&+b_{(0,0),(0,1)}+b_{(0,0),(1,1)}.
\end{split}
\end{align}

It is straightforward to check that all of these anyons are bosons. We can check that they are closed under fusion using the $S$ matrix elements defined in (\ref{smatrix}) and the Verlinde formula (\ref{verlinde}).

Out of all the $a_g^{(0)},a_g^{(1)},$ and $b_{g,h}$, we have the following bosons (aside from the ones in $\mathcal{L}_e$): $a_{(1,1)}^{(0)},a_{(1,1)}^{(1)},$ and $b_{(1,0),(0,1)}$. We will now compute the self statistics of the anyons $c_g^{(0)}$ and $c_g^{(1)}$. From the previous section, we have
\begin{align}
\begin{split}
\Omega^{(1,0),\sigma\sigma\sigma}_{c_{(0,0)}^{(0)}}&=\Omega^{(0,1),\sigma\sigma\sigma}_{c_{(0,0)}^{(0)}}=i\\
\Omega^{(1,0),\sigma\sigma\sigma}_{c_{(1,0)}^{(0)}}&=-\Omega^{(0,1),\sigma\sigma\sigma}_{c_{(1,0)}^{(0)}}=-i\\
\Omega^{(1,0),\sigma\sigma\sigma}_{c_{(0,1)}^{(0)}}&=-\Omega^{(0,1),\sigma\sigma\sigma}_{c_{(0,1)}^{(0)}}=i\\
\Omega^{(1,0),\sigma\sigma\sigma}_{c_{(1,1)}^{(0)}}&=\Omega^{(0,1),\sigma\sigma\sigma}_{c_{(1,1)}^{(0)}}=-i.
\end{split}
\end{align}

This gives
\begin{align}
\begin{split}
e^{i\theta_{c_{(0,0)}^{(0)}}}&=\sqrt{i\epsilon}\qquad e^{i\theta_{c_{(1,0)}^{(0)}}}=\sqrt{\epsilon}\\
 e^{i\theta_{c_{(0,1)}^{(0)}}}&=\sqrt{\epsilon}\qquad e^{i\theta_{c_{(1,1)}^{(0)}}}=\sqrt{-i\epsilon},
\end{split}
\end{align}
and $e^{i\theta_{c_g^{(1)}}}=-e^{i\theta_{c_g^{(0)}}}$.

We see that for $\epsilon=1$, $c_{(1,0)}^{(0)}$ and $c_{(0,1)}^{(0)}$ have bosonic statistics. We propose that there is a gapped, symmetric, $(1+1)D$ theory from the following $\mathcal{L}_m$:
\begin{equation}
\mathcal{L}_m=a_{(0,0)}^{(0)}+a_{(1,1)}^{(1)}+b_{(1,0),(0,1)}+c_{(1,0)}^{(0)}+c_{(0,1)}^{(0)}.
\end{equation}
Note that we cannot confirm that this forms a valid Lagrangian algebra without solving for the $M$ symbol. 

For $\epsilon=-1$, none of the $c_g^{(0)},c_g^{(1)}$ anyons have bosonic self statistics. We therefore do not have enough anyons to form a $\mathcal{L}_m$; the fusion category symmetry is anomalous.

We now proceed to the off-diagonal bicharacter, given by (\ref{offdiagf}). $\mathcal{L}_e$ is the same as in the diagonal case. We now have
\begin{align}
\begin{split}
\Omega^{(1,0),\sigma\sigma\sigma}_{c_{(0,0)}^{(0)}}&=\Omega^{(0,1),\sigma\sigma\sigma}_{c_{(0,0)}^{(0)}}=1\\
\Omega^{(1,0),\sigma\sigma\sigma}_{c_{(1,0)}^{(0)}}&=-\Omega^{(0,1),\sigma\sigma\sigma}_{c_{(1,0)}^{(0)}}=-1\\
\Omega^{(1,0),\sigma\sigma\sigma}_{c_{(0,1)}^{(0)}}&=-\Omega^{(0,1),\sigma\sigma\sigma}_{c_{(0,1)}^{(0)}}=1\\
\Omega^{(1,0),\sigma\sigma\sigma}_{c_{(1,1)}^{(0)}}&=\Omega^{(0,1),\sigma\sigma\sigma}_{c_{(1,1)}^{(0)}}=-1.
\end{split}
\end{align}

This gives
\begin{align}
\begin{split}
e^{i\theta_{c_{(0,0)}^{(0)}}}&=\sqrt{\epsilon}\qquad e^{i\theta_{c_{(1,0)}^{(0)}}}=\sqrt{\epsilon}\\
 e^{i\theta_{c_{(0,1)}^{(0)}}}&=\sqrt{\epsilon}\qquad e^{i\theta_{c_{(1,1)}^{(0)}}}=i\sqrt{\epsilon},
\end{split}
\end{align}
and $e^{i\theta_{c_g^{(0)}}}=-e^{i\theta_{c_g^{(1)}}}$. We see that for $\epsilon=1$, $c_{(0,0)}^{(0)},c_{(1,0)}^{(0)},$ and $c_{(0,1)}^{(0)}$ all have bosonic statistics. We propose the following three magnetic Lagrangian algebras:
\begin{align}
\begin{split}\label{loffdiag}
\mathcal{L}_{m_1}&=a_{(0,0)}^{(0)}+a_{(1,0)}^{(1)}+a_{(0,1)}^{(0)}+a_{(1,1)}^{(0)}+2c_{(1,0)}^{(0)}\\
\mathcal{L}_{m_2}&=a_{(0,0)}^{(0)}+a_{(1,0)}^{(0)}+a_{(0,1)}^{(1)}+a_{(1,1)}^{(0)}+2c_{(0,1)}^{(0)}\\
\mathcal{L}_{m_3}&=a_{(0,0)}^{(0)}+a_{(1,0)}^{(0)}+a_{(0,1)}^{(0)}+a_{(1,1)}^{(1)}+2c_{(0,0)}^{(0)}.
\end{split}
\end{align}

By computing the $S$-matrix via (\ref{smatrix}), we checked that these three Lagrangian algebras satisfy a set of necessary, but not sufficient, conditions for $\mathcal{L}$ being a Lagrangian algebra \cite{lan2015,kaidi2022higher}:
\begin{equation}\label{stcond}
SZ_{\mathcal{L}}=Z_{\mathcal{L}}\qquad TZ_{\mathcal{L}}=Z_{\mathcal{L}}.
\end{equation}
 Here, $Z_{\mathcal{L}}$ is a $|\mathcal{C}|$-component vector with entries $\{n_a\}$. 
This agrees with the result that there are three $(1+1)D$ SPTs of this fusion category symmetry\cite{tambara1998, bhardwaj2018,thorngrenwang1}. 

For $\epsilon=-1$, $c_{(1,1)}^{(1)}$ is a boson, so we propose a single $\mathcal{L}_m$ given by
\begin{equation}
\mathcal{L}_m=a_{(0,0)}^{(0)}+a_{(1,0)}^{(1)}+a_{(0,1)}^{(1)}+a_{(1,1)}^{(1)}+2c_{(1,1)}^{(1)}.
\end{equation}

\subsection{Example: $G=\mathbb{Z}_ N\times\mathbb{Z}_N$ for $N>2,N$ odd}\label{sexample}
For $N$ odd, (\ref{omegapi}) simplifies to
\begin{equation}
\left(\Omega^{(1,0),\sigma\sigma\sigma}_{c_g^{(r)}}\right)^N=\left(\Omega^{(0,1),\sigma\sigma\sigma}_{c_g^{(r)}}\right)^N=1.
\end{equation}

We choose to label
\begin{equation}
\Omega^{(1,0),\sigma\sigma\sigma}_{c_g^{(r)}}=e^{-\frac{2\pi ip}{N}}\qquad\Omega^{(0,1),\sigma\sigma\sigma}_{c_g^{(r)}}=e^{-\frac{2\pi iq}{N}},
\end{equation}
for $p,q\in[0,N-1]$. The $N^2$ different elements $g=(p,q)$ correspond to the $N^2$ different duality anyons in $\mathcal{Z}[\mathrm{TY}_{\mathbb{Z}_N\times\mathbb{Z}_N}^{\chi,\epsilon}]$. 

Using the notation of (\ref{bicharacterxyz}) and the result in (\ref{result2}), we obtain
\begin{equation}
\Omega^{(n,m),\sigma\sigma\sigma}_{c_g^{(r)}}=e^{-\frac{2\pi ipn}{N}}e^{-\frac{2\pi iqm}{N}}x^{\frac{n(n-1)}{2}}y^{\frac{m(m-1)}{2}}z^{nm}
\end{equation}

We can plug this into (\ref{nonabelianstat}) to get the topological spins of the duality anyons for any choice of $x,y,z$. To determine the number of bosonic duality anyons for $\epsilon=+1$, we need to determine the number of distinct pairs $(p,q)$ that give a solution to
\begin{equation}
\frac{1}{N}\sum_{n,m}\Omega^{(n,m),\sigma\sigma\sigma}_{c_g^{(r)}}=1
\end{equation}
For $\epsilon=-1$, we need to find the number of pairs that give a solution to
to 
\begin{equation}
\frac{1}{N}\sum_{n,m}\Omega^{(n,m),\sigma\sigma\sigma}_{c_g^{(r)}}=-1
\end{equation}

Let us determine the number of bosonic duality anyons for $\epsilon=\pm1$ for the diagonal and off-diagonal bicharacters. The diagonal bicharacter corresponds to $x=y=e^{\frac{2\pi i}{N}},z=1$. In this case, we show in Appendix~\ref{ssum} that when $-1$ is a quadratic residue mod $N$, there are $2N-1$ bosonic duality anyons when $\epsilon=+1$ and zero bosonic duality anyons when $\epsilon=-1$. The off-diagonal bicharacter corresponds to $x=y=1,z=e^{\frac{2\pi i}{N}}$. In this case, we obtain
\begin{equation}
\frac{1}{N}\sum_{n,m}\Omega^{(n,m),\sigma\sigma\sigma}_{c_g^{(r)}}=e^{-\frac{2\pi ipq}{N}}
\end{equation}

We immediately see that there are $2N-1$ solutions to $e^{-\frac{2\pi ipq}{N}}=+1$, given by $p=0$ or $q=0$, and no solutions to $e^{-\frac{2\pi ipq}{N}}=-1$ for any odd $N$. Therefore, there are $2N-1$ bosonic duality anyons for $\epsilon=+1$ and none for $\epsilon=-1$.

\section{Bosonic duality anyons for the diagonal bicharacter}\label{ssum}
For the diagonal bicharacter, we have
\begin{equation}\label{rss1diagapp}
\sum_{n,m}\Omega^{(n,m),\sigma\sigma\sigma}_{c_g^{(r)}}=\sum_{n,m=0}^{N-1}e^{\frac{i\pi n(n-1-2p)}{N}}e^{\frac{i\pi m(m-1-2q)}{N}}.
\end{equation}
We would like to find the pairs $(p,q)$ that give 
\begin{equation}\label{req1}
\frac{1}{N}\sum_{n,m=0}^{N-1}e^{\frac{i\pi n(n-1-2p)}{N}}e^{\frac{i\pi m(m-1-2q)}{N}}=1,
\end{equation}
and we would like to show that there does not exist any pair $(p,q)$ for which
\begin{equation}\label{req2}
\frac{1}{N}\sum_{n,m=0}^{N-1}e^{\frac{i\pi n(n-1-2p)}{N}}e^{\frac{i\pi m(m-1-2q)}{N}}=-1.
\end{equation}

We use the quadratic reciprocity law (notice that the requirements for using this are satisfied because $N$ is odd) for general Gauss sums\cite{berndt1998} to obtain
\begin{equation}
\sum_{n=0}^{N-1}e^{\frac{i\pi n(n-1-2p)}{N}}=\sqrt{N}e^{\frac{i\pi (N-(1+2p)^2)}{N}},
\end{equation}
which gives
\begin{align}
\begin{split}
&\frac{1}{N}\sum_{n,m=0}^{N-1}e^{\frac{i\pi n(n-1-2p)}{N}}e^{\frac{i\pi m(m-1-2q)}{N}}\\
&=ie^{-\frac{i\pi}{2N}(1+2p+2p^2+2q+2q^2)},
\end{split}
\end{align}
so (\ref{req1}) means we need to search for solutions to 
\begin{equation}\label{modcond}
1+2p(p+1)+2q(q+1)=N\qquad\text{mod }4N.
\end{equation}

Note that choosing $p=q=\frac{N-1}{2}$ is always a solution. Plugging this in, we get
\begin{equation}\label{Npyth}
N^2=N\qquad\text{mod }4N\to N=1\qquad{\text{mod }}4,
\end{equation}
which is always satisfied when $-1$ is a quadratic residue mod $N$. Specifically, for the prime factors of these $N$ must all be pythagorean, leading to (\ref{Npyth}). To find the other solutions, let $p=\frac{N-1}{2}+\delta_1$ and $q=\frac{N-1}{2}+\delta_2$. The previous solution corresponds to $\delta_1=\delta_2=0$. Plugging this into (\ref{modcond}), we get
\begin{align}
\begin{split}
&1+\left(N-1+2\delta_1\right)\left(\frac{N+1}{2}+\delta_1\right)\\
&+\left(N-1+2\delta_2\right)\left(\frac{N+1}{2}+\delta_2\right)=N\qquad\text{mod }4N.
\end{split}
\end{align}
Pulling out the $N^2$ part from the previous solution, we get
\begin{equation}\label{2Ndelta}
2N\delta_1+2N\delta_2+2\delta_1^2+2\delta_2^2=0\qquad\text{mod }4N.
\end{equation}
Now let $\delta_1$ be any integer in the range $\left[-\frac{N-1}{2},\frac{N-1}{2}\right]$, and let $\delta_2=s\delta_1$. Plugging this into (\ref{2Ndelta}) and dividing through by 4, we get
\begin{equation}
N\delta_1(1+s)+\delta_1^2(1+s^2)=0\qquad\text{mod }N.
\end{equation}

Because $-1$ is a quadratic residue mod $N$, we can always choose $s\in[0,N-1]$ such that $s^2=-1$ mod $N$, which clearly gives a solution for any $\delta_1$. In fact there are always only two solutions: $s$ and $N-s$, giving $q=\frac{N-1}{2}+s\delta_1$ and $q=\frac{N-1}{2}-s\delta_1$. So we see that for each of the $N$ values of $\delta_1$ there are two solutions, except at $\delta_1=\delta_2=0$. As a result we have a total of $2N-1$ solutions to (\ref{modcond}), giving the $2N-1$ bosonic duality defects.

For (\ref{req2}), we need to show that there are no $(p,q)$ satisfying
\begin{equation}
1+2p(p+1)+2q(q+1)=N(4n+3).
\end{equation}

Let $N=4k+1$ where $k$ is an integer. Then we have
\begin{align}
\begin{split}
1+2p(p+1)+2q(q+1)&=(4k+1)(4n+3)\\
2p(p+1)+2q(q+1)&=16kn+12k+4n+2\\
p(p+1)+q(q+1)&=8kn+6k+2n+1.
\end{split}
\end{align}

The left hand side must be even because either $p$ or $p+1$ is even, and either $q$ or $q+1$ is even. However, the right hand side is clearly odd. Therefore there is no solution to this equation.

\section{Duality-invariant Lagrangian subgroups and the hyperbolic bicharacter condition}\label{shyperbolic}
We will show in this appendix that the condition that there exists a duality-invariant $\tilde{\mathcal{L}}_m$ of the $G$ gauge theory is actually equivalent to the condition that the bicharacter is hyperbolic, as defined in (\ref{hyperbolic1}) and (\ref{hyperbolic}). First, we observe that if a $G$ gauge theory has a duality-invariant $\tilde{\mathcal{L}}_m$, then we can generate the $|G|^2$ abelian anyons of the gauge theory by 
\begin{equation}
\mathcal{C}=\tilde{\mathcal{L}}_{m}^{(0)}\times\tilde{\mathcal{L}}_{m}^{(1)}
\end{equation}
where $\tilde{\mathcal{L}}_{m}^{(0)}$ and $\tilde{\mathcal{L}}_{m}^{(1)}$ satisfy the following properties. First, they are each are mapped back to themselves under the duality symmetry, because they are duality invariant. Second, the anyons in each subgroup are all bosonic (but anyons formed by the fusion of one in each subgroup are not bosonic in general). Third, they are each, as groups, are isomorphic to $G$. This comes from the fact that at least one of the two Lagrangian subgroups, say $\tilde{\mathcal{L}}_{m}^{(0)}$, must be magnetic. This means that it must contain an anyon from each equivalence class under modding out by gauge charges (see Sec.~\ref{sabelian}), and must be closed under fusion. Therefore it must, as a group, be isomorphic to $G$. This is violated in $\mathbb{Z}_N$ gauge theory when $N$ is a perfect square, where the (not magnetic) duality invariant $\tilde{\mathcal{L}}$ is isomorphic to $\mathbb{Z}_{\sqrt{N}}\times\mathbb{Z}_{\sqrt{N}}$ rather than $\mathbb{Z}_N$. Since $\tilde{\mathcal{L}}_m^{(0)}\cong G$, in order to generate all of the $G\times G$ anyons, $\tilde{\mathcal{L}}_{m}^{(1)}$ must also be isomorphic to $G$ as a group.

A general property of any duality-invariant group of anyons is that it can be written as $\mathcal{C}=\mathcal{C}_1\times\mathcal{C}_2$, where $\mathcal{C}_1$ and $\mathcal{C}_2$ are mapped onto each other under duality. When $\mathcal{C}$ is braided, this result comes from the anyon content of the gauged theory, which is the Drinfeld center of a TY fusion category (see Sec.~\ref{sgaugeduality} and Appendix~\ref{sstringnet}). Specifically, it comes from the number of abelian anyons vs dimension two anyons in the Drinfeld center of a TY fusion category. An alternative intuitive argument is the following: first, we cannot have every anyon in $\mathcal{C}$ be invariant under duality; the duality symmetry would then have trivial action. Now suppose that a single anyon $a$ of $\mathcal{C}$ is mapped to a different anyon $a_d$ under the duality symmetry, so that it forms an orbit of size two under the duality symmetry. Then fusing $a$ with any other anyon $b\in\mathcal{C}$ will also give an an orbit of size two. $a\times b$ is not invariant under duality even if $b$ is invariant under duality. Therefore, it is not possible to have $|\mathcal{C}|-2$ duality invariant anyons and a single orbit of size two. Continuing in this way, using the nature of the abelian anyon fusion, one finds that $\mathcal{C}$ must take the form $\mathcal{C}=\mathcal{C}_1\times\mathcal{C}_2$, where $\mathcal{C}_1$ and $\mathcal{C}_2$ are mapped onto each other under duality.

In particular, this means that
\begin{align}
\begin{split}
\tilde{\mathcal{L}}_{m}^{(0)}&=\tilde{\mathcal{L}}_{m1}^{(0)}\times \tilde{\mathcal{L}}_{m2}^{(0)}\\
\tilde{\mathcal{L}}_{m}^{(1)}&=\tilde{\mathcal{L}}_{m1}^{(1)}\times \tilde{\mathcal{L}}_{m2}^{(1)}
\end{split}
\end{align}
where $\tilde{\mathcal{L}}_{m1}^{(0)}$ and $ \tilde{\mathcal{L}}_{m2}^{(0)}$ are mapped onto each other under duality, as are $\tilde{\mathcal{L}}_{m1}^{(1)}$ and $ \tilde{\mathcal{L}}_{m2}^{(1)}$. $\tilde{\mathcal{L}}_{m1}^{(0)},\tilde{\mathcal{L}}_{m2}^{(0)},\tilde{\mathcal{L}}_{m1}^{(1)},$ and $\tilde{\mathcal{L}}_{m2}^{(1)}$ are all isomorphic to $G_1\cong G_2$ as groups. Notice that combining an anyon in $\tilde{\mathcal{L}}_{m1}^{(0)}$ with its duality partner in $\tilde{\mathcal{L}}_{m2}^{(0)}$ gives a bosonic duality invariant anyon. The $\sqrt{|G|}$ bosonic duality invariant anyons in $\tilde{\mathcal{L}}_m^{(0)}$ form a group isomorphic to $G_1$, and the $\sqrt{|G|}$ bosonic duality invariant anyons in $\tilde{\mathcal{L}}_m^{(1)}$ also form a group isomorphic to $G_1$, giving $2\sqrt{|G|}-1$ duality invariant bosons (where we subtracted 1 to avoid double counting the vacuum anyon). The fact that these duality invariant anyons are bosonic (which come from the fact that they belong in Lagrangian subgroups) is precisely the condition that $\chi(g_1,g_1')=\chi(g_2,g_2')=1$, and in particular $\chi(g_1,g_1)=\chi(g_2,g_2)=1$. Specifically, $\chi(g,h)$ is given by the topological spins of the anyons in $G$ gauge theory according to (\ref{topspinab}), and $\chi(g,g)$ gives the topological spins of the duality-invariant anyons. The hyperbolic condition (\ref{hyperbolic}) guarantees that there are at least $2\sqrt{|G|}-1$ bosonic duality invariant anyons in the $G$ gauge theory, which matches the result above from the existence of a duality-invariant $\tilde{\mathcal{L}}_m$.

\let\oldaddcontentsline\addcontentsline
\renewcommand{\addcontentsline}[3]{}
\bibliography{tybib}
\let\addcontentsline\oldaddcontentsline

\end{document}